
\overfullrule=0pt
\input amstex
\documentstyle{vanilla}
%
%
\font\tenbf=cmbx10

\font\ninebf=cmbx9
\font\ninerm=cmr9
\font\nineit=cmti9

\font\bigbf=cmbx10 scaled \magstep1
\font\bigrm=cmr10  scaled \magstep1
\font\bigmit=cmmi10 scaled \magstep1
\def\title{\begingroup\baselineskip=16pt
 \textfont0=\bigrm\textfont1=\bigmit\bigbf}
\let\endtitle=\endgroup
\TagsOnRight
\let\ms=\mathsurround
\def\sectiontitle#1[]{\vskip0pt plus.1\vsize\penalty-250
 \vskip0pt plus-.1\vsize\bigskip\vskip\parskip
 \message{SectionTitle}\leftline{\tenbf#1}\nobreak\vglue 5pt}
\def\qed{\hbox{${\vcenter{\vbox{
 \hrule height 0.4pt\hbox{\vrule width 0.4pt height 6pt
 \kern5pt\vrule width 0.4pt}\hrule height 0.4pt}}}$}}
\def\MyP#1{\medbreak\noindent\smc\ignorespaces#1\unskip.\enspace\sl
 \ignorespaces}
\let\EndP=\endproclaim
\def\gad#1{\global\advance#1 by1}
\newcount\Sno
\Sno=0
\def\ST#1/ST{\gad\Sno\Fno=0\lmno=0\Dno=0\sectiontitle{\the\Sno. #1}[]}
\def\STE#1/ST{\gad\Sno\Fno=0\lmno=0\Dno=0
 \sectiontitle{\hbox to\parindent{\hfil}#1}[]}
\newcount\lmno
\lmno=0
\outer\def\Lm#1{\gad\lmno\xdef#1{\the\Sno.\the\lmno}\MyP{Lemma #1}}
\newcount\Dno
\Dno=0
\outer\def\Df{\gad\Dno\MyP{Definition \the\Sno.\the\Dno}}
\outer\def\Pf#1.{\par\ifdim\lastskip<\smallskipamount\removelastskip
 \smallskip\fi\noindent{\smc\ignorespaces Proof #1\unskip.\enspace}
 \rm\ignorespaces}
\outer\def\EPf{\qed\par\smallskip}
\outer\def\Cor{\par\ifdim\lastskip<\smallskipamount\removelastskip
 \smallskip\fi\noindent{\ignorespaces{\it Corollary.\/}\enspace}
 \rm\ignorespaces}
\outer\def\EndC{\par\smallskip}
\newcount\Fno
\Fno=0
\def\FTg#1{\ifx#1\undefined\gad\Fno\xdef#1{\the\Sno.\the\Fno}\fi\tag{#1}}
\newcount\Rno
\Rno=0
\def\Rf#1{\ifx#1\undefined\gad\Rno\xdef#1{\the\Rno}\fi#1}
\def\Rfh#1{\ifx#1\undefined\gad\Rno\xdef#1{\the\Rno}\fi}
\def\ns{\ms=0pt}
\let\nin=\noindent
\let\cl=\centerline
\let\st=\subset
\let\vp=\vphantom

\let\x=\otimes
\def\xa{\x_\A}
\def\xc{\x_\C}
\let\der=\partial
\let\lto=\longrightarrow
\let\lr=\longleftrightarrow
\def\({\allowbreak(}
\let\alb=\allowbreak
\let\amb=\allowmathbreak
\let\adb=\allowdisplaybreaks

\let\fsp=\frenchspacing
\def\Bm{\vp{\bigg|}}
\def\vk#1>{\vbox to #1{\vss}}
\def\>{\relax\ifmmode\mskip.25\thinmuskip\relax\else\kern.04167em\fi}
\def\<{\relax\ifmmode\mskip-.25\thinmuskip\relax\else\kern-.04167em\fi}
\def\nn#1>{\noalign{\vskip #1}}
\def\mv#1#{\kern#1\move#1[]}
\def\move#1[]#2{#2\kern-#1}
\def\rit#1{\itemitem{\hbox to 15pt{(#1)\hss}}}
\def\ov#1#2{\overset{\>\>{#1}\<\<}\to{#2}}
\def\un#1#2{\underset{#1}\to{#2}}
\def\bcup{\bigcup\limits}
\def\prodl{\prod\limits}
\def\prodi{\prodl^{\vp{i}}}
\def\prodT#1#2{{\ns\mathchoice
 {\hbox{$\dsize\kern-7pt\mathop{\kern7pt\prod\nolimits^
  {\lower2.3pt\hbox{$\ssize\x$}}}_{#1}^{#2}$}}
 {\mathop{{\prod}^{\lower1.7pt\hbox{$\ssize\x$}}}
  \limits_{\mkern-16mu #1}^{\mkern-15mu #2}}{nechto}{nechto}}}
\def\prodt#1{{\prodT{#1=1}{nM}}}
\def\diag{{\text{\rm diag}\,}}
\def\detq{\det\nolimits_q}
\def\crk{\text{\rm corank}\,}
\def\tr{{\text{\rm tr}\,}}
\def\I{{\text{\rm im}\,}}
\def\E{{\text{\rm End}\,}}
\def\ld{\,\ldots}
\def\cld{\cdot\ldots\cdot}
\def\lld{\land\ldots\land}
\def\xld{\x\ldots\x}
\def\qc{\quad,}
\def\qp{\quad.}
\let\qq=\qquad
\def\A{{\Cal A}}
\def\H{{\Cal H}}
\def\J{{\Cal J}}
\def\K{{\Cal K}}
\def\Lc{{\Cal L}}
\def\M{{\Cal M}}
\def\O{{\Cal O}}
\def\Pc{{\Cal P}}
\def\Q{{\Cal Q}}
\def\Rc{{\Cal R}}
\def\Sc{{\Cal S}}
\def\T{{\Cal T}}
\def\U{{\Cal U}}
\def\Vc{{\Cal V}}
\def\Wc{{\Cal W}}
\def\X{{\Cal X}}
\def\Z{{\Cal Z}}
\def\1{\text{\bf 1}}
\def\am{{\bold a}}
\def\bb{{\bold b}}
\def\F{{\bold F}}
\def\G{{\bold G}}
\def\ibb{{\bold i}}
\def\jbb{{\bold j}}
\def\kbb{{\bold k}}
\def\lbb{{\bold l}}
\def\p{{\bold p}}
\def\Pb{{\bold P}}
\def\R{{\bold R}}
\def\s{{\bold s}}

\def\vb{{\bold v}}
\def\GG{{\Cal G}}
\def\L{{\text{\rm L}}}
\def\Sr{{\text{\rm S}}}
\def\C{{\bold C}}
\def\Cn{{\bold C}^n}
\def\Ah{{\widehat A}}
\def\Bh{{\widehat B}}
\def\Ch{{\widehat C}}
\def\Dh{{\widehat D}}
\def\Hh{{\widehat H}}
\def\hh{{\widehat h}}
\def\a{{\tilde a}}
\def\be{{\tilde b}}
\def\pbe{{\tilde\p}}
\def\pe{{\tilde p}}
\def\pee{{\hat p}}
\def\Tt{{\tilde T}}

\def\B{{\bar b}}
\def\lb{{\bar\ell}}
\def\db{{\bar d}}
\def\Rb{{\ns\mathchoice
 {\ov{\vbox to.5pt{\vfil\smash{$\ssize\>-\<$}}}{\vbox to4.3pt
  {\vfil\smash{$R$}}}}
 {\ov{\vbox to.5pt{\vfil\smash{$\ssize\>-\<$}}}{\vbox to4.3pt
  {\vfil\smash{$R$}}}}
 {\ov{\vbox to0pt{\smash{$\sssize\mskip.18\thinmuskip-\mskip-.18\thinmuskip$}}}
  {\vbox to2.33pt{\vfil\smash{$\ssize R$}}}}
 {nichto}}}
\def\Psb{{\ns\mathchoice
 {\ov{\vbox to.5pt{\vfil\smash{$\ssize\mv-.47pt{-}$}}}{\vbox to4.3pt
  {\vfil\smash{$\Psi$}}}}
 {\ov{\vbox to.5pt{\vfil\smash{$\ssize\mv-.47pt{-}$}}}{\vbox to4.3pt
  {\vfil\smash{$\Psi$}}}}
 {\bar\Psi}{\bar\Psi}}}
\def\al#1#2{\alpha_{#1#2}}
\def\bt#1#2{\beta_{#1#2}}
\def\gm#1#2{\gamma_{#1#2}}
\def\dl#1#2{\delta_{#1#2}}
\let\eps=\epsilon
\def\e#1#2{{\eps_{#1#2}}}

\def\beps{\bar\eps}
\def\eb#1#2{{\beps_{#1#2}}}
\let\veps=\varepsilon
\let\la=\lambda
\def\om#1#2{\omega_{#1#2}}
\def\omb#1#2{\bar\omega_{#1#2}}
\let\si=\sigma
\let\Si=\Sigma
\def\rhb{{\bar\rho}}
\def\chb{{\bar\chi}}
\let\Ups=\Upsilon
\def\UM{\Ups_M}
\def\zt#1#2{\zeta_{#1#2}}
\def\ab#1{\Phi_{#1#1}}

\def\abt{a_i\be_i}
\def\bat{b_i\a_i}
\def\abN{a_i^N\be_i^N}
\def\baN{b_i^N\a_i^N}
\def\Aif{A^\infty}
\def\Ao{{\ov{\ \circ}\A}}
\def\AM{{\ov{\ \circ}\A_\Sigma}}
\def\FG#1{F_{#1+1}\x G_{#1}}
\def\ij{_{\>\ibb\>\jbb}}
\def\kiN{{\kappa_i^N}}
\def\Ld{\Lc^\dagger}
\def\Lb{\Ld_m}
\def\Lo{\Lc^\circ}
\def\Lp{\Lc_m(\p)}
\def\Ml{\M_\T[v]}
\def\NM{{N^{(n-1)nM/2}}}
\def\oh#1{{\widehat\omega_{#1}}}
\def\omt#1{\omega^{\eta_{#1}}}
\def\phip{\phi_m(\p)}
\def\pia{\pi(\Ao)}
\def\pip{\pi_m(\p)}
\def\pii{{\pi_1,\pi_2}}
\def\pit{{\pi_1\x\pi_2}}
\def\pti{{\pi_2\x\pi_1}}
\def\po{p^\circ}
\def\pp{(p,\pe)}
\def\ppp{(p,\pe,\pee)}
\def\RA{\R_{12}(\pii)}
\def\RB{\R_{13}(\pi_1,\pi_3)}
\def\RC{\R_{23}(\pi_2,\pi_3)}
\def\Rep{\Re(\p_1,\p_2,\p_3)}
\def\Rk{{\ov{\!\rlap{$\sssize k,k+1$}\phantom{k,k}}R(\omega)}}
\def\Rbk{{\ov{\!\rlap{$\sssize k,k+1$}\phantom{k,k}}\Rb(\omega)}}
\def\so{\si^\circ}
\def\Td{{\detq T(u)}}
\def\Te#1;{T\w #1;}
\def\tij{{\theta_{ij}}}
\def\Tl{\T_\la}
\def\Tm{T^{\x_qm}(u)}
\def\TN{T^{\x_qN}(u)}
\def\TNh{\hat\TN}
\def\Tov#1{\ov{#1}T\w(n-1);}
\def\V#1;{V\w #1;}
\def\VA#1;{\V{#1};_\A}
\def\W{\ov\circ{\hbox{\vbox to5pt{}\ns$\smash\Wc$}}}
\def\Wm{\W\tm}
\def\Wo{\Wc^\circ}
\def\av#1{{\langle{#1}\rangle}}
\def\sav{^{\sssize \av{}}}
\def\saT{^{\sssize \av T}}
\def\map{{\ \buildrel\phi\over\longrightarrow\ }}
\def\w#1;{^{\wedge{#1}}}
\def\ot#1;{^{\x #1}}
\def\tm{\ot m;}
\def\vst#1{{\lower2.1pt\hbox{$\bigr|_#1$}}}

\def\id,{\,,\amb\ \,i=1,\ld,}
\def\il,{i=1,\ld,}

\def\usM{(\si_1,\ld,\si_{nM})}
\def\Uq{U_q(\widehat{gl}(n))}
\def\Ucc#1{\uppercase{#1}}
\def\Uc#1/{\edef\Uz{\noexpand\Ucc#1/}\Uz}
\def\point#1;#2;#3;#4;{\pmatrix #1&\ldots&#2\\#3&\ldots&#4\endpmatrix}
\def\points#1;#2;#3;#4;#5;#6;{\pmatrix #1&\ldots&#2&#3\\#4&\ldots&#5&#6
 \endpmatrix}
\def\SA/{section 2}
\def\SB/{section 8}
\def\SC/{section 6}
\def\SD/{section 7}
\def\Rm/{{\ns$R$}-matrix}
\def\Rms/{{\ns$R$}-matrices}
\def\YB/{Yang-Baxter equation}
\def\mm/{monodromy matrices}
\def\alg/{the algebra}
\def\amm/{algebra of monodromy matrices}
\def\fnd/{finite-\dm/al}
\def\rp/{rep\-res\-en\-ta\-tion}
\def\eqt/{equivalent}
\def\ir/{irreducible}
\def\irp/{\ir/ \rp/}
\def\ccm/{cocommuting}
\def\crp/{\ccm/ \rp/s}
\def\pln/{po\-ly\-nom\-ial}
\def\prp/{\pln/ \rp/}
\def\Apn/{{\ns$A$}-\pln/}
\def\Arp/{{\ns$A$}-\rp/}
\def\Airp/{\ir/ \Arp/}
\def\frp/{elementary \rp/}
\def\srp/{simplest \rp/}
\def\tp/{tensor product}
\def\sp/{spectral parameter}
\def\Bw/{Boltzmann weight}
\def\cPm/{chiral Potts model}
\def\ipf/{interpolating formulae}
\def\rcp/{recursive process}
\def\itw/{intertwiner}
\def\gnl/{generically}
\def\cel/{central element}
\def\prl/{principal}
\def\qm/{quantum minor}
\def\qdet/{quantum determinant}
\def\rl/{relation}
\def\crl/{commutation relations}
\def\fpr/{fusion procedure}
\def\Lop/{{\ns$L$}-operator}
\def\cz/{commutant}
\def\sub/{the subalgebra}
\def\inv/{invariant}
\def\isp/{\inv/ subspace}
\def\irs/{\ir/ subspace}
\def\fcrp/{factorizable \rp/}
\def\wrt/{with respect to}
\def\gb/{generated by}
\def\hm/{homomorphism}
\def\crd/{completely reducible}
\def\dm/{dim\-en\-sion}
\def\inr/{inversion relation}
\def\evl/{eigenvalue}
\def\ev/{eigenvector}
\def\cev/{common \ev/}
\def\esp/{eigenspace}
\def\mcs/{maximal common \esp/}
\def\irc/{\ir/ component}
\def\vv/{vice versa}
\def\cor/{corollary}
\def\crr/{correspond}
\def\cnv/{convenient}
\def\Am/{{\ns$\A$-module}}
\def\Abm/{{\ns$\A$}-bimodule}
\def\sbm/{submodule}
\def\ism/{\inv/ \sbm/}
\def\tri/{trigonometric}
\hsize=4.6truein
\vsize=7truein
\footline={\ifnum\pageno=1 \hfil \else
 \ifodd\pageno \ninerm\hfil\folio \else \ninerm\folio\hfil \fi\fi}
\baselineskip=13pt
\line{\hfil\ninerm hep-th/9211105}
\kern-\baselineskip
\baselineskip=12pt
\kern.4truein
\cl{\sl RIMS -- 903}
\kern1truein
\title
\cl{Cyclic \mm/}
\cl{for $sl(n)$ \tri/ \Rms/}
\endtitle
\kern.6truein
\cl{Vitaly TARASOV}
\kern.32truein
\cl{\it Physics Department, Leningrad University}
\cl{\it Leningrad 198904 Russia}
\kern.16truein
\cl{\it Research Institute for Mathematical Sciences}
\cl{\it Kyoto University, Kyoto 606 Japan}
\kern1truein
\nin{\bf Abstract.}\enspace\, The \amm/ for $sl(n)$ \tri/ \Rm/ is studied.
It is shown that a generic \fnd/ \pln/ \irp/ of this algebra is \eqt/ to a \tp/
of \Lop/s. Cocommutativity of \rp/s is discussed and \itw/s for \fcrp/s are
written through the \Bw/s of the  $sl(n)$ \cPm/.
\vfil\break
\baselineskip=13pt
\parindent=15pt
\ms=1.6pt
%
\Sno=-1
\STE Introduction/ST
Let us consider an algebra \gb/ noncommutative entries of the
matrix $T(u)$ satisfying the famous bilinear \rl/ originated from the quantum
inverse scattering method [\Rf\TF]
$$
R(\la-\mu)T(\la)T(\mu)=T(\mu)T(\la)R(\la-\mu)
$$
where $R(\la)$ is \Rm/ -- a solution of the \YB/. For historical
reasons this algebra is called the \amm/. It possesses a natural
bialgebra structure with the coproduct (1.5). If $\GG$ is a simple \fnd/ Lie
algebra and $R(\la)$ is $\GG$-invariant \Rm/ the \amm/ after
a proper specialization gives the Yangian $Y(\GG)$ introduced by
Drinfeld [\Rf\DRI]. If $R(\la)$ is \crr/ing \tri/
\Rm/ [\Rf\BAZH,\Rf\JMA] (see (1.1) for $sl(n)$ case) this algebra is closely
connected  with $U_q(\GG)$ and $U_q(\widehat\GG)$ at zero level
[\DRI,\JMA--\Rfh\JM \Rf\RTF]. In the last case it is \cnv/ to
use new variable $u=\exp\la$ rather than $\la$.
If $R(\la)$ is $sl(2)$ elliptic \Rm/ [\Rf\BAX,\Rf\BEL] the \amm/ gives rise to
Sklyanin's algebra [\Rf\SKLY].
\par
In this paper we shall study algebras of \mm/ for $sl(n)$ \tri/
{\ns$R$}-matrices [\Rf\CHER]. In the framework of the quantum inverse
scattering method \fnd/ \irp/s of these
algebras which depend polynomially on the \sp/ $u$ are
of the special interest. They \crr/ to integrable models on a
finite lattice. \Lop/s are \irp/s with
linear dependence on the \sp/, and usually we get
\prp/ as a \tp/ of \Lop/s. The
question is to examine whether all \fnd/ polynomial \irp/s
can be obtained in this way. For the $sl(2)$ case
\crr/ing to the \Rm/ of the six-vertex model the answer is
known. If $\omega$ is generic that each wanted \rp/ is
\eqt/ to a \tp/ of \Lop/s [\Rf\TAR,\Rf\TARJ]. If $\omega$
is a root of 1 the situation is more complicated. In this case only
generic \rp/ are \eqt/ to \tp/s of \Lop/s, but there are also exist \rp/s,
which are not of this form [\TARJ]. For generic $\omega$ in the $sl(n)$ case
\fnd/ \irp/s was described in [\Rf\DRII,\Rf\CHERD], but to obtain all of them
from \Lop/s the notion of an \Lop/ should be generalized.
Here we study $sl(n)$ case for $\omega$ being  a root of 1 and
obtain the same results as for the $sl(2)$ case [\TARJ].
\par
As well known, the deformation parameter being a root of 1 is a
peculiar case for quantum groups [\Rf\KdC]. It is the same for algebras of
\mm/ under consideration if $\omega^N=1$. In this case a generic
polynomial \fnd/ \irp/ is cyclic (without highest and lowest
vectors). Moreover, as usual \irp/s do not cocommute, their tensor
products in direct and inverse orders are not \eqt/ in contrast
to what takes place for generic $\omega$. The whole set of \irp/s
exfoliate to varieties of \crp/. For a couple of \crp/ one can define
an \itw/ realizing an equivalence of two \tp/s. Intertwiners
gives us solutions of the \YB/, \rp/s playing a role of \sp/s.
In the $sl(2)$ case an \itw/ for \Lop/s can be written as a
product of four factors and each of them can be expressed explicitly through
the \Bw/s of the \cPm/ [\TARJ,\Rf\BS]. A direct generalization of this
construction for the $sl(n)$ case leads to the $sl(n)$ \cPm/ [\Rf\BMKS]
and minimal \rp/s of $\Uq$ [\Rf\JMM]. Unfortunately, minimal
\Lop/s from [\BMKS] (which \crr/ to minimal
\rp/s of $\Uq$ [\JMM]) are not generic from the point of view
of this paper. For a generic \Lop/ necessary factorization if exists
contains $n$ factors instead of two factors for minimal one, so an
\itw/ is a product of $n^2$ factors. But explicit expressions
for these factors can be written through the same \Bw/ of the $sl(n)$ \cPm/.
Recently, another factorization for a generic \Lop/ was obtained and \crr/ing
 formula for an \itw/ was written by use of the same \Bw/ [\Rf\KAMA].
\par
The paper is organized as follows. In the first section we give
definitions and formulate results without proofs. Next two sections
contain proofs of the Theorems 1,2\,. In the forth section we introduce
factorized \Lop/s and build their \itw/s; the connection
with the $sl(n)$ \cPm/ is also discussed. In the last sections we give
technical details and necessary proofs. Some proofs which can be done by
explicit calculation are omitted.
\ST The \amm//ST
Let us define an \amm/ for the $sl(n)$ \tri/ \Rm/. Denote for short
$\M=\E\Cn$. The \Rm/ $R(u)$ is considered as an element of $\M\ot2;$ and has
the following nonzero entries:
$$
\aligned
R_{ii}^{ii}(u)&=1-u\omega \qc\\
R_{ij}^{ij}(u)&=\om ij(1-u)\qc\qq
R_{ji}^{ij}(u)=u^\tij (1-\omega )\qc\qq i\ne j
\endaligned
\FTg\fa
$$
where $\tij=\left\{\matrix1\ ,&i<j\\0\ ,&i\ge j\endmatrix
\right.,\ \om ij\om ji=\omega^{1+\dl ij}$ and $\dl ij$ is the Kronecker symbol.
We also introduce a tensor $\eps$ such that $\om ij=\omega^\e ij$.
This definition of $R(u)$ differs slightly from the original one [\CHER].
A variable $u$ is called the \sp/. $R(u)$  satisfies the \YB/:
$$
\ov{12}R(u)\ov{13}R(uv)\ov{23}R(v)=\ov{23}R(v)
\ov{13}R(uv)\ov{12}R(u)\qp
$$
Here we use the standard matrix notations, superscripts indicating the way
of embeddings $\M\st\M\ot3;$ as \crr/ing factors.
\Df
The \amm/ $\A$ is the
algebra defined by generators $T_{ij}(u),\,H_i\ \,i,j=1,\ld,n$
and \rl/s
$$
\gather
R(u)\ov1T(uv)\ov2T(v)=\ov2T(v)\ov1T(uv)R(u)
\FTg\fca\\
\nn3pt>
\left[\oh l\x H_l\,,\,T(u)\right]=0\qc\qq
\oh l=\diag(1,\ld,\un{l\text{\rm-th}}{\omega},\ld,1)\qc\qq
\FTg\fcb\\
\nn1pt>
H_iH_j=H_jH_i\qc\qq\prod_lH_l=1
\endgather
$$
where $T(u)\in\M\x\A$ with entries $T_{ij}(u)\in\A$.
\EndP
\nin Here and later $\prodl_l\equiv\prodl_{l=1}^n$ and the
same convention is implied for sums. More explicit form Eq.(\fcb) is
$$
H_lT_{ij}(u)=T_{ij}(u)H_l\omega^{\dl lj-\dl li}\qp
\FTg\fcc
$$
One can introduce the natural coproduct $\Delta:\,\A\to\A\ot2;$:
$$
\aligned
\Delta(T(u))&=T_1(u)T_2(u)\in\M\x\A\ot2;\qc\\
\Delta(H_l)&=H_l\x H_l\\
\endaligned
\FTg\fcd
$$
(subscripts indicate the way of embedding $\A\st\A\ot2;$)
and counit $\veps:\A\to\C$:
$$
\veps(T(u))=I\qc\qq\veps(H_l)=1
$$
making $\A$ a bialgebra, hence a \tp/ of $\A$-modules is also $\A$-module.
\Uc\alg/ $\A$ is closely connected with \alg/ $\Uq$,
but does not exactly coincide with it. In the \SB/ we shall discuss the
structure of \alg/ $\A$ in more details.
\par
We are interesting in a special class of \rp/s of \alg/ $\A$. Often the \rp/
will be indicated by a superscript.
\Df
A \rp/ $\pi$ of \alg/ $\A$ is called a \prp/ if $\dim\pi<\infty$, $T^{\pi}(u)$
is polynomial on $u$ and $T^{\pi}_{ij}(0)=0$ for $i<j$. \alb
$\deg\pi\equiv\deg T^{\pi}=\un{ij}{\max}\,(\tij+\deg T^{\pi}_{ji})$ is
called a degree of the \rp/ $\pi$ .
\EndP
\Uc\alg/ $\A$ has the well known element $\detq T(u)$ which is called the
\qdet/ (the exact definition of $\detq T(u)$ is given in \SC/).
Henceforward we assume that all $\e ij$ are integer.
\Lm\LA
$\dsize Q(u)=\detq T(u)\,\prod_{il}H_l^\e li$
is a \cel/.
\EndP
\Pf. In \SC/. \EPf
\Lm\LE
$\Delta(\detq T(u))=\detq T(u)\x\detq T(u)$
\EndP
\Pf. In \SD/. \EPf
\nin For a \prp/ $\pi\,,\,\deg\pi=M,\,T(u)\equiv T^{\pi}(u)$ we define
$$
\align
T_{ii}(u)&=T_{ii}^{\infty}(-u)^M+\ldots+T_{ii}^0\qc\\
\nn3pt>
T_{ij}(u)&=(-u)^\tij(T_{ij}^{\infty}(-u)^{M-1}+\ldots+T_{ij}^0)\qc
\!\!\qq i\ne j\<\qc
\FTg\fe\\
\nn3pt>
Q(u)&=Q^{\infty}(-u)^{nM}+\ld+Q^0\qp
\endalign
$$
\Lm\LB
Let $\pi$ be a \prp/,
$T(u)\equiv T^{\pi}(u)\,,H_i\equiv H_i^{\pi}$. Operators
$t_i^{\infty}=T_{ii}^{\infty}\cdot\prodl_lH_l^{-\e il}$ and
$t_i^0=T_{ii}^0\cdot\prodl_lH_l^\e li$ commute with
$T(u)\,,H_1,\ld,H_n$.
\EndP
\nin It is obvious that
$Q^{\infty}=\prodl_it_i^{\infty}$ , $Q^0=\prodl_it_i^0$.
\par
Henceforth throughout the paper we take $\omega$ being a primitive $N$-th
root of 1\,. In this case \alg/ $\A$ has
an additional large set of \cel/s. To describe them explicitly we introduce
an operation $\av\cdot$ as follows: $\av\O(u^N)=\prodl_{k=1}^N\O(u\omega^k)$.
\Lm\LC
$\av{T_{ij}}(v)\,,H_1^N,\ld,H_n^N$ are \cel/s.
\EndP
\Pf. In \SD/. \EPf
\nin Define the element $\av T(v)\in\M\x\A$ such that
$\av T_{ij}(v)=\av{T_{ij}}(v)$.
\Lm\LF
$\Delta(\av T(v))=\av{T_1}(v)\av{T_2}(v)$.
\EndP
\Pf. In \SD/.\EPf
\Df
 For any $\T\in\M$ let $A_k^\T\,,B_k^\T\,,C_k^\T$ be
the following minors:
\item{} $A_k^\T$ is the \prl/ minor \gb/ the first $k$
rows and columns.
\item{} $B_k^\T$ is \gb/ the first $k$ rows and
$k+1$ columns (except the $k$-th column).
\item{} $C_k^\T$ is \gb/ the first $k+1$ rows and
$k$ columns (except the $k$-th row).
\EndP
\Df
$\T(v)\in\M[v]\,,\,\deg\T=M$ is called an \Apn/ if it enjoys the properties
\rit 1 $\T_{ij}(0)=0\,\ $ if $i<j$.
\rit 2 $\deg\T_{ij}<M\,\ $ if $i>j$.
\rit 3 For any $k<n\ \ A_k^\T(v)$ has exactly $kM$ nonzero simple zeros.
\rit 4 If $A_k^\T(v_0)=0$ then $B_k^\T(v_0)\ne0$ and $C_k^\T(v_0)\ne0$.
\par\nin $A\M[v]$ denotes the set of all \Apn/s.
\EndP
\edef\DA{\the\Sno.\the\Dno}
\nin It is evident that $\deg A_k^\T=kM,\,\deg B_k^\T\le kM,\,
\deg C_k^\T<kM$ and $A_k^\T(0)\ne0\,,\amb B_k^\T(0)=0$.
\Df
$\UM$ is a variety of sets
$\Si=\{\T(v)\in A\M[v]\,,\Q(u)\in\C[u]\,,h_i\,,\amb z_i^\infty\,,\amb z_i^0\}
_{i=1}^n$ such that $\deg\T=M$ and
$$
\adb
\gather
\T_{ii}(v)=\left((-v)^M(z_i^\infty)^N+\ldots+(z_i^0)^Nh_l^{-1}\right)
\prod_lh_l^\e il\qc\\
\nn1pt>
\Q(u)=(-u)^{nM}\prod_iz_i^\infty\,+\ldots+\prod_iz_i^0\qc\\
\nn2pt>
\det\T(v)=\av\Q(v)\prod_{il}h_l^\e il\qc\qq\prod_lh_l=1\qp
\FTg\ff
\endgather
$$
\EndP
\Lm\LCA
$\UM$ is diffeomorphic to a dense open set in $\C^{n^2M+2n-1}$.
\EndP
\Pf. In \SA/.\EPf
\Df
The \prp/ $\pi$ is called an \Arp/ if $\av T^\pi(v)\in A\M[v]$ and
$\deg\av T^\pi=\deg\pi$. An \Airp/ of degree 1 is called an \frp/ (\Lop/).
\EndP
\nin For any \Airp/ $\pi$ we put
$$
\Si^\pi=\{\av T^\pi(v)\,,Q^\pi(u)\,,(H_l^N)^\pi\,,(t_i^\infty)^\pi\,,
(t_i^0)^\pi\}\qp
$$
\Lm\LCB
$\Si^\pi\in\UM\,,\ M=\deg\pi$.
\EndP
\Pf. In \SD/.\EPf
\proclaim{Theorem 1} For any set $\Si\in\UM$ there exists a unique \Airp/
$\pi$ such that $\Si^\pi=\Si$. Moreover, $\deg\pi=M$ and $\dim\pi=\NM$.
\EndP
\proclaim{Theorem 2} A generic \Airp/ of degree $M\ge1$ is \eqt/ to a \tp/ of
$M$ \frp/s.
\EndP
\nin{\it Note.\/} One can check if a \rp/ $\pi$ is \eqt/ to a \tp/ of \frp/s
using only $\av{T}^\pi(v)$.
\ST The Proof of Theorem 1. Uniqueness/ST
In order to prove the Theorem 1 we shall describe the construction of an \Airp/
inspired by Drinfeld's new realization of Yangians [\DRII] and the ideas of the
functional Bethe ansatz [\Rf\SKL]. Let us introduce the
special elements of \alg/ $\A$ -- \qm/s of $T(u)$; the exact definition and
the calculation of \crl/ for \qm/s is given in \SC/. The following \qm/s will
play an important role:
\item{} $\Ah_k(u)$ is a \prl/ minor \gb/ the first $k$
rows and columns;
\item{} $\Bh_k(u)$ is \gb/ the first $k$ rows and $k+1$ columns (except the
$k$-th column);
\item{} $\Ch_k(u)$ is \gb/ the first $k+1$ rows and $k$ columns (except the
$k$-th row);
\item{} $\Dh_k(u)$ is \gb/ the first $k+1$ rows and columns (except the
$k$-th row and column);
\par
\nin It is also \cnv/ to introduce improved minors whose
\crl/ are simpler than for original ones:
$$
\align
&{\aligned
A_k(u)&=\Ah_k(u)\Hh_k\ \ ,\ \quad B_k(u)=\Bh_k(u)\Hh_k\qc\\
C_k(u)&=\Ch_k(u)\Hh_k\ \ ,\ \quad
D_k(u)=\Dh_k(u)\Hh_{k-1}\prod_lH_l^{-\e {k+1,}l}\rlap{\ \,,}
\endaligned}
\FTg\gaa\\
\nn-6pt>
&\,\Hh_k=\prod_{i=1}^k\prod_lH_l^{-\e il}\ \ .
\endalign
$$
Main \crl/ read as follows:
$$
\adb
\gather
\gathered
[A_i(u)\,,A_j(v)]=[A_i(u)\,,H_l]=0\\
\nn2pt>
[A_i(u)\,,B_j(v)]=[A_i(u)\,,C_j(v)]=[B_i(u)\,,C_j(v)]=0
\ \,,\quad i\ne j\ ,\\
\nn2pt>
[B_i(u)\,,B_i(v)]=[C_i(u)\,,C_i(v)]=0\qc
\endgathered
\FTg\gba\\
\nn4pt>
H_lB_i(u)=\omega^{\dl {i+1,}l-\dl il}B_i(u)H_l\qc\qq
H_lC_i(u)=\omega^{\dl il-\dl {i+1,}l}C_i(u)H_l\qc\\
\nn5pt>
\aligned
B_i(u)B_j(v)&=\omega^{\eta_{ji}}B_j(v)B_i(u)\\
C_i(u)C_j(v)&=\omega^{\eta_{ij}}C_j(v)C_i(u)
\endaligned
\qc\qq |i-j|>1\qc
\FTg\gbb\\
\nn3pt>
\eta_{ij}=\e i{,j+1}+\e{i+1,}j-\e ij-\e{i+1,}{j+1}\hskip.85in\\
\nn9pt>
\aligned
(u-v)A_i(u)B_i(v)&=(u-v\omega)B_i(v)A_i(u)-v(1-\omega)B_i(u)A_i(v)\\
\omega(u-v)A_i(u)C_i(v)&=(u\omega-v)C_i(v)A_i(u)+u(1-\omega)C_i(u)A_i(v)
\endaligned
\FTg\gbc\\
\nn8pt>
D_i(u)A_i(u\omega)-\omega B_i(u)C_i(u\omega)H^{(i)}=
A_{i+1}(u\omega)A_{i-1}(u)\qc
\FTg\gbd\\
\nn9pt>
H^{(i)}=\prod_lH_l^{\e il-\e{i+1,}l}
\FTg\gc
\endgather
$$
where $A_0(u)=1\ ,\ A_n(u)=Q(u)$. Note that
$$
H^{(i)}B_j(u)=\omega^{\eta_{ij}}B_j(u)H^{(i)}\qc\qq
H^{(i)}C_j(u)=\omega^{-\eta_{ij}}C_j(u)H^{(i)}\qp
$$
\par
\nin Let us also define improved minors of $\av T(v)$:
$$
\aligned
A\sav_k(v)&=A\saT_k(v)\Hh^N_k\qc\qq
B\sav_k(v)=B\saT_k(v)\Hh^N_k\qc\\
C\sav_k(v)&=C\saT_k(v)\Hh^N_k
\endaligned
\FTg\gca
$$
\nin where minors $A\saT_k(v)\,,B\saT_k(v)\,,C\saT_k(v)$
were defined above.
\Lm\LD
$\av{A_i}(v)=A\sav_i(v)\ ,\quad\av{B_i}(v)=B\sav_i(v)\ ,\quad
\av {C_i}(v)=C\sav_i(v)\ .$
\EndP
\Pf. In \SD/. \EPf
\nin Denote by $\Ao$ \sub/ \gb/
$\{\Ah_k(u)\,,\Bh_k(u)\,,\Ch_k(u)\,,H_k\}_{k=1}^{n-1}$.
Certainly, $\Ao$ is also \gb/ $\{A_k(u)\,,B_k(u)\,,C_k(u)\,,H_k\}_{k=1}^{n-1}$.
\par
Now let us fix throughout this section an \Airp/ $\pi$ of degree $M$ and take
all elements of \alg/ $\A$ in this \rp/. (The explicit indication of $\pi$ will
be omitted.) Let $\{\zt ij\}$ be the set of all zeros of the \pln/
$A\sav_i(v)$. Because $\pi$ is an \Arp/, all these zeros are nonzero and
simple. Introduce operators $\al kj\,,\bt kj\,,\gm kj$ as follows
$$
\gather
A_k(u)=\Aif_k\prod^{kM}_{j=1}(\al kj-u)\ \ ,\quad\ \al kj^N=\zt kj\ \ ,\ \quad
\Aif_k=\prod^k_{i=1}t_i^\infty\ \ ,
\FTg\gda\\
\nn3pt>
\bt ij=B_i(\al ij)\qc\qq\gm ij=C_i(\al ij)\qp
\FTg\gdb
\endgather
$$
When substituting $\al ij$ instead of the \sp/ the ordering of
noncommuting factors has to be chosen. We prefer to put all $\alpha$'s to the
right, but one can choose another ordering and all the following remains
correct. Eq.(\gba)--\(\gc) and the Lemma {\LD} lead to the following
\rl/s for these operators:
$$
\adb
\gather
\gathered
[\al ik\,,\al jl]=[\al ik\,,H_l]=[H_i\,,H_l]=0\qc\\
\nn3pt>
\al ik\bt jl=\bt jl\al ij\omega^{\dl ij\dl kl}\qc\qq
\al ik\gm jl=\gm jl\al ij\omega^{-\dl ij\dl kl}\qc\\
\nn4pt>
H_i\bt jl=\omega^{\dl i{,j+1}-\dl ij}\bt jlH_i\qc\qq
H_i\gm jl=\omega^{\dl ij-\dl i{,j+1}}\gm jlH_i\qc
\endgathered
\FTg\gga\\
\nn4pt>
\gathered
[\bt ik\,,\bt il]=[\bt ik\,,\gm jl]=[\gm ik\,,\gm il]=0\qc\ \ \quad
i\ne j\!\qc\\
\nn2pt>
\mv-30pt{\bt ik\bt jl=\bt jl\bt ik\omega^{\eta_{ij}}\ \ ,\ \quad
\gm ik\gm jl=\gm jl\gm ik\omega^{-\eta_{ij}}\ \ ,
\ \quad\rlap{\ns$|i-j|>1\!\qc$}}
\endgathered
\FTg\ggaa\\
\nn3pt>
\aligned
\omega\bt ik\gm ikH^{(i)}&=-A_{i+1}(\al ik)A_{i-1}(\al ik\omega^{-1})\qc\\
\nn2pt>
\gm ik\bt ikH^{(i)}&=-A_{i+1}(\al ik\omega)A_{i-1}(\al ik)\qc
\endaligned
\FTg\ggb\\
\nn3pt>
\bt ij^N=B\sav_i(\zt ij)\qc\qq\gm ij^N=C\sav_i(\zt ij)\qc
\FTg\ggab\\
\nn3pt>
\Aif_k\prod^{kM}_{j=1}\al kj=\prod^k_{i=1}t^0_iH_i^{-1}\qp
\FTg\ggc
\endgather
$$
Since $\pi$ is an \Arp/ $\bt ij$ and $\gm ij$ are invertible (see (\ggab)).
 For present the definition (\gda) of operators $\al ij$ is formal. To make it
sensible we introduce a vector $\vb$ -- a \cev/ of $A_i(u)\id,n-1$ and the
subspace $V=\pia\vb$.
\Lm\LDA
\item{1.}$V$ is spanned by \cev/s of $A_i(u)$ with different \evl/s.
\item{2.}$\al ij\,,\bt ij\,,\gm ij$ can be well defined on $V$ as operators
satisfying \rl/s \(\gga)--\(\ggc).
\item{3.}$\dim V=\NM$.
\EndP
\Pf ({\it Sketch\/}). Evidently we can define $\al ij$ on $\vb$ claiming $\vb$
to be its \ev/ with the appropriate \evl/. Then the subspace $V$ can be set up
step by step starting from $\vb$ by use of $\bt kl$ and $\gm kl$. At every step
the definition of $\al ij$ can be naturally extended to fulfil \rl/s \(\gga).
It is easy to check that this construction can be realized self-consistently
giving the subspace $V$ of the required \dm/ and operators
$\al ij\,,\bt ij\,,\gm ij$ on it satisfying the \rl/s (\gga)--\(\ggc).
And for the operators $B_k(u)$ and $C_k(u)$ we have the \ipf/:
$$
\gather
B_k(u)=u\sum_{i=1}^{kM}\bt ki\al ki^{-1}P_{ki}(u)\!\qc\!\!\qq
C_k(u)=\sum_{i=1}^M\gm kiP_{ki}(u)
\FTg\gh\\
\nn3pt>
\line{where\hfil$\dsize P_{ki}(u)=\prod_{\ssize{j=1}\atop\ssize{j\ne i}}^{kM}
{{u-\al kj}\over{\al ki-\al kj}}\qp$\hfil}
\endgather
$$
\EPf
\nin{\it Note.\/}
By the definition of {\ns$\alpha$'s} one can retell the first point saying that
$V$ is spanned by \cev/s of {\ns$\alpha$'s} with different \evl/s.\newline
One can also see that for $\vb'$ -- another \cev/ of $A_i(u)$ \ $V$ and
$V'=\pia\vb'$ are isomorphic as {\ns$\pia$}-orbits.
\par
To complete this part of the proof of the Theorem 1 it is enough to show that
$V$ is invariant \wrt/ $\pi(\A)$.  To have more compact notations we shall
show that $\pi(\A)\st\pia$ using $\al ij\,,\bt ij\,,\gm ij$.
The way of doing this is the following \rcp/.
The first step is trivial:
$$
T_{11}(u)=\Ah_1(u)\qc\qq T_{12}(u)=\Bh_1(u)\qc\qq T_{21}(u)=\Ch_1(u)
$$
(see (\gaa),\(\gc)). $T_{22}(u)$ can be tested by means of the \rl/
$$
\Ah_2(u\omega)=T_{22}(u\omega)T_{11}(u)-\om 21T_{21}(u\omega)T_{12}(u)\qp
$$
To pass to the 3~by~3 \prl/ submatix one has to use \rl/s
$$
\aligned
\Bh_2(u\omega)&=\om 21\left(\om 31^{-1}T_{23}(u\omega)T_{11}(u)-
T_{21}(u\omega)T_{13}(u)\right)\qc\\
\nn3pt>
\Ch_2(u\omega)&=\om 31\left(\om 21^{-1}T_{32}(u\omega)T_{11}(u)-
T_{31}(u\omega)T_{12}(u)\right)\qp
\endaligned
\FTg\gk
$$
Substituting here $u=\al 1i$ we obtain the \ipf/ for
$T_{13}(u)\,,\amb T_{31}(u)$:
$$
\aligned
T_{13}(u)&=-u\om21^{-1}\Hh_1\sum_{i=1}^M\gm 1i^{-1}\Bh_2(\al 1i\omega)
\al 1i^{-1}P_{1i}(u)\qc\\
T_{31}(u)&=-\om 32^{-1}\Hh_1\sum_{i=1}^M\Ch_2(\al 1i\omega)\bt 1i^{-1}
P_{1i}(u)\qp
\endaligned
$$
Now $T_{23}(u)\,,T_{32}(u)\in\pia$ due to (\gk) and to test $T_{33}(u)$ we
recall that
$$
\Bm\Ah_3(u\omega)=T_{33}(u\omega)\Ah_2(u)+\,\text{\it known terms}\qp
$$
\par
\nin For further steps we have to introduce additional \qm/s:
\item{} $\Bh_{kl}(u)$ is \gb/ the first $k$ rows and $k-1$ columns together
with {\ns$(k+l)$}-th column;
\item{} $\Ch_{kl}(u)$ is \gb/ the first $k-1$ rows and $k$ columns together
with {\ns$(k+l)$}-th row;
\item{} $\Dh^B_{kl}(u)$ is \gb/ the first $k-1$ rows and columns together with
{\ns$(k+l)$-th row and $(k+1)$-th column};
\item{} $\Dh^C_{kl}(u)$ is \gb/ the first $k-1$ rows and columns together with
{\ns$(k+1)$-th row and $(k+l)$-th column}.
\par
\nin We also define the \crr/ing improved minors:
\par\kern-13pt
$$
\aligned
B_{kl}(u)&=\Bh_{kl}(u)\Hh_k\qc\qq
D^B_{kl}(u)=\Dh^B_{kl}(u)\Hh_{k-1}\prod_lH_i^{-\e {k+1,}i}\\
C_{kl}(u)&=\Ch_{kl}(u)\Hh_k\qc\qq
D^C_{kl}(u)=\Dh^C_{kl}(u)\Hh_{k-1}\prod_lH_i^{-\e {k+1,}i}
\endaligned
$$
\par\kern-10pt\nin
(cf.(\gaa)) and use the \rl/s
\par\kern-15pt
$$
\kern-5pt
\aligned
\Bm D^B_{kl}(u)A_k(u\omega)-\omega B_{kl}(u)C_k(u\omega)H^{(k)}&=
{\om{k+l,}k\over\om{k+1,}k}B_{k+1,l-1}(u\omega)A_{k-1}(u)\\
D^C_{kl}(u)A_k(u\omega)-\omega B_k(u)C_{kl}(u\omega)H^{(k)}&=
{\om{k+1,}k\over\om{k+l,}k}C_{k+1,l-1}(u\omega)A_{k-1}(u)
\endaligned
\FTg\gna
$$
\par\kern-5pt\nin
which look similar to (\gbd). To check $T_{i4}(u)\in\pia\,,\ \,i=1,2,3$ the
following formulae have to be written:
\par\kern-22pt
$$
\gather
H^{(2)}B_{22}(u)=-u\om34^{-1}\sum_{i=1}^{2M}\gm 2i^{-1}
B_3(\al 2i\omega)A_1(\al 2i)\al 2i^{-1}P_{2i}(u)\qc
\FTg\gnb\\
\nn4pt>
\Bh_{22}(u\omega)=\om 21\left(\om 41^{-1}T_{24}(u\omega)T_{11}(u)-
T_{21}(u\omega)T_{14}(u)\right)\qc
\FTg\gnc\\
\nn4pt>
T_{14}(u)=-u\om21^{-1}\Hh_1^{-1}\sum_{i=1}^M\gm 1i^{-1}\Bh_{22}(\al 1i\omega)
\al 1i^{-1}P_{1i}(u)\qc
\FTg\gnd
\endgather
$$
\par\kern-5pt\nin
Eq.(\gnb),\(\gnd) are obtained from the first of Eq.(\gna) for $k=2$
and Eq.(\gnc) respectively after the following substitutions:
$u=\al 2i\omega^{-1}$ and $u=\al 1i$. Now $T_{24}(u)\in\pia$ due to (\gnc) and
to test $T_{34}(u)$ we use
$$
\om41\om42\Bh_3(u\omega)=\om31\om32T_{34}(u\omega)\Ah_2(u)
+\text{\it known terms}\qp
$$
In the same manner we can show that
$T_{4i}(u)\in\pia\,,\ \,i=1,2,3$. In order to test $T_{44}(u)$ and thus to
complete this step of the process we look to
$$
\Ah_4(u\omega)=T_{44}(u\omega)\Ah_3(u)+\text{\it known terms}\qp
$$
It is quite evident how to do next steps by means of \rl/s (\gna) and \ipf/.
As a result of this \rcp/ we can express all $T_{kl}(u)$ through operators
$\al ij\,,\bt ij\,,\gm ij$. Justifying this formal calculations like in the
Lemma {\LDA} we convince ourselves that $\pi(\A)V\st V$.
\Pf of Lemma \LCA.
The \rcp/ desribed above certainly has the ``classical limit'' -- a very
similar one for usual matrix \pln/s. It shows that the variety $\UM$ can be
parametrized by $\Q(u)$, minors $A_i^\T(v)\,,B_i^\T(v)\,,C_i^\T(v)\id,{n-1}$
and $h_i\,,z_i^\infty\,,z_i^0\id,n$. Now it is very easy to find independent
parameters in which the identity mapping is the required diffeomorphism.
\EPf
\ST The Proof of Theorem~1. Existence/ST
Let a set $\Si\in\UM$ be given. We have to find an \Airp/ $\pi$ such that
$\Si=\Si^\pi$. Define \alg/ $\AM$ by generators
${\{\al ik,\bt ik,\gm ik,H_i\}_{i=1}^M}\vp\}_{\,k=1}^{\,iM}$
and \rl/s (cf.(\gca)--\(\ggc)):
$$
\adb
\gather
\gathered
[\al ik\,,\al jl]=[\al ik\,,H_l]=[H_i\,,H_l]=0\qc\\
\nn3pt>
\al ik\bt jl=\bt jl\al ij\omega^{\dl ij\dl kl}\qc\qq
\al ik\gm jl=\gm jl\al ij\omega^{-\dl ij\dl kl}\qc\\
\nn4pt>
H_i\bt jl=\omega^{\dl i{,j+1}-\dl ij}\bt jlH_i\qc\qq
H_i\gm jl=\omega^{\dl ij-\dl i{,j+1}}\gm jlH_i\qc
\endgathered\\
\nn4pt>
\gathered
[\bt ik\,,\bt il]=[\bt ik\,,\gm jl]=[\gm ik\,,\gm il]=0\qc\ \ \quad
i\ne j\!\qc\\
\nn2pt>
\mv-30pt{\bt ik\bt jl=\bt jl\bt ik\omega^{\eta_{ij}}\ \ ,\ \quad
\gm ik\gm jl=\gm jl\gm ik\omega^{-\eta_{ij}}\ \ ,
\ \quad\rlap{\ns$|i-j|>1\!\qc$}}
\endgathered\\
\nn3pt>
\aligned
\omega\bt ik\gm ikH^{(i)}&=-A_{i+1}(\al ik)A_{i-1}(\al ik\omega^{-1})\qc\\
\nn2pt>
\gm ik\bt ikH^{(i)}&=-A_{i+1}(\al ik\omega)A_{i-1}(\al ik)\qc
\endaligned\\
\nn3pt>
\al ij^N=\zt ij\qc\qq
\bt ij^N=\hh_iB^\T_i(\zt ij)\qc\qq\gm ij^N=\hh_iC^\T_i(\zt ij)\qc\\
\nn3pt>
\prod_{i=1}^kz_i^\infty\prod^{kM}_{j=1}\al kj=\prod^k_{i=1}z^0_iH_i^{-1}\qc\\
\nn3pt>
A_k(u)=\prod_{i=1}^kz_i^\infty\prod^{kM}_{j=1}(\al kj-u)\qc\qq
H^{(i)}=\prod_lH_l^{\e il-\e{i+1,}l}\qc\\
\hh_k=\prod_{i=1}^k\prod_lh_l^{-\e il}\qp
\endgather
$$
It easy to see that $\AM$ is a simple algebra isomorphic to
$\E\C^{\NM}$ so it has a unique \irp/ and any its \rp/ is faithful.
Before we have shown that an \Airp/ $\pi$ generates the \irp/
of \alg/ $\Ao_{\Si^\pi}$. Now we would like to reverse a logic. Let
$B(u)\,,C(u)$ be defined by Eq.(\gh) and $\Ah(u)\,,\Bh(u)\,,\Ch(u)$ by
Eq.(\gaa). Define the \hm/ $\varphi:\A\to\AM$ on generators as follows:
$\varphi(H_i)=H_i$ and $\varphi(T_{ij}(u))$ is given by the \rcp/ described
in the previous section. For the definition of $\varphi$ to be correct
all the \rl/s (\fa) have to be preserved by $\varphi$. To verify this is to
check some \pln/ identities on $\UM$. So they have to be
checked only for generic $\Si$ and it certainly will be done if an \Airp/
$\pi$ such that $\Si^\pi=\Si$ will be shown. Though we return almost to the
starting point of the consideration we have a profit to solve the problem only
for generic $\Si$. In this case the required \Airp/ can be built from some
simple primitives.
\par
Later we shall treat $\Cn$-coordinate indices modulo $n$, excepting the cases
when they appear in inequalities. Introduce \alg/ $\Wc$ \gb/
$F_i\,,G_i\,,H_i\id,n$ and \rl/s
$$
\gather
 F_iF_j=F_jF_i\qc\qq F_iH_j=H_jF_i\qc\qq H_iH_j=H_jH_i\qc\\
\om ijF_iG_j=G_jF_i\om i{,j+1}\qc\qq
H_iG_j=G_jH_i\omega^{\dl i{,j+1}-\dl ij}\qc
\FTg\gpa\\
\nn3pt>
\om ij G_iG_j=G_jG_i\om {i+1,}{j+1}\qc\qq\prod_lH_l=1\qp
\endgather
$$
Let $f_i=F_i\prodl_lH_l^{-\e il}\!,\,\F=F_1\ldots F_n$, and $\G=G_1\ldots G_n$.
Elements $f_i\,,\amb F_i^N\!,\amb\,G_i^N\!,\amb\,H_i^N\!\!\id,n$
and $\F\G^{-1}$ clearly
generate the center of $\Wc$. The mapping $\phi:\A\to\Wc$:
$$
\aligned
T_{ij}(u)&\map -uF_i\dl ij+(-u)^\tij G_i\dl{i+1,}j\qc\\
H_l&\map H_l
\endaligned
\FTg\gqa
$$
is a \hm/ of algebras. It is easy to calculate that
$$
\aligned
Q(u)&\map (-u)^{n-1}\omega^{(1-n)n/2}
\Bigl((-1)^n\G\prod_{i=2}^n\om 1i-u\F\Bigr)\qc\\
\nn6pt>
\av{T_{ij}}(v)&\map -vF_i^N\dl ij+(-v)^\tij G_i^N\dl {i+1,}j\qp
\endaligned
$$
 For any \rp/ $\xi$ of \alg/ $\Wc$ the \rp/ $\xi\circ\phi$ of \alg/ $\A$
will be called a \srp/.
\par
Let $\Vc=\E\C^N$ and $X,Z\in\Vc$ be the following matrices:
$X_{ij}=\dl i{,j+1\!\pmod N\,},\amb Z_{ij}=\omega^i\dl ij$.
Define naturally operators $X_i\,,Z_i\in\Vc\ot n;$:
$$
X_i=I\ot(i-1);\x X\x I\ot(n-i-1);\qc\qq
Z_i=I\ot(i-1);\x Z\x I\ot(n-i-1);
$$
and introduce the subspace $\H\st(\C^N)\ot n;$ as the \esp/
$Z\ot n;=1$.
\EndP
\Lm\LFC
Let $a_i\,,b_i\,,c_i\id,n$ be arbitrary numbers such that $\prodl_ic_i=1$
and $m_{ij}\,,\ \,i,j=1\ld,n$ be integers such that
$m_{i,l+1}-m_{il}-m_{l,i+1}-m_{li}=\e{i+1,}{l+1}-\e il$. The mapping
$\xi:\Wc\to\E\H$:
\par\kern-15pt
$$
 F_i=a_i\prodi_lZ_l^{\e il}\ \ ,\ \quad
G_i=b_iX_{i+1}X_i^{-1}\prod_lZ_l^{m_{il}}\ \ ,\ \quad
H_i=c_iZ_i
$$
\par\kern-4pt\nin
is a representation of \alg/ $\Wc$.
\EndP
Now we have got a lot of \srp/s to extract the required \Airp/ from a
\tp/ of \srp/s. Let $\kappa_i\id,nM$ be zeros of $\Q(u)$
(simple for generic case), and let us take nonzero vectors
$\Psi_i\in\ker\T(\kiN)$ which are unique up to scale factor due to (\ff).
Define step by step a sequence of \srp/s $\si_i=\xi_i\circ\phi$ such that
$$
\gather
\gathered
Q^{\si_i}(\kappa_i)=0\qc\qq\av T^{\si_i}(\kiN)\Psi_{ii}=0\qc\\
\nn2pt>
\prodt j\xi_j(f_l)=z_l^\infty\qc\qq
\prodt j\si_j(H_l^N)=h_l\qc
\endgathered
\FTg\gsa\\
\nn3pt>
\line{where\hfil$\Psi_{1i}=\Psi_i\qc\qq
\Psi_{i+1,j}=\av T^{\si_i}(\kappa_j^N)\Psi_{ij}$\hfil}
\endgather
$$
and take the \rp/ $\pi=\pi\usM$ such that
$$
\aligned
T^\pi(u)&=(-u)^{(1-n)M}T_{nM}(u;\si_{nM})\cld T_1(u;\si_1)\qc\\
\nn2pt>
H^\pi_i&=\prodt j\si_j(H_i)\qc\qq T_i(u;\si_i)=T_i^{\si_i}(u)\qp
\endaligned
\FTg\gsb
$$
\Lm\LG
$\av T^\pi(v)=\T(v)$.
\EndP
\Pf.
Consider the ratio $\tau(v)=\av T^\pi(v)\T^{-1}(v)$. This is a meromorphic
function having poles only at points $\kiN$. But Eq.(\gsa),\(\gsb) show that
for any $i\ \,\un{\,n=\kiN}{\text{res}\,}\tau(u)=0$. Hence, $\tau(v)$ does not
depend on $v$. Taking limits $v\to 0$ and $v\to\infty$ we see that $\tau(v)$
is both an upper triangular matrix with unit diagonal and a lower triangular
one. Then $\tau(v)$ is the unit matrix.
\EPf
One can easily check that the \rp/ $\pi$ is a \pln/ \Arp/ of degree $M$ and
$Q^\pi(u)=\Q(u)\,,\,t_i^\infty=z_i^\infty$.
\vadjust{\kern1pt}
As a \cor/ of the Lemma {\LG} we have got that
$(t_i^0)^N\!=\prodt j\si_j(G_{i-j}^N)=(z_i^0)^N\!$. According to the
Lemma {\LB} the \rp/ $\pi$ can be restricted to a \mcs/ of operators
$t_i^0=\prodt j\xi_j\bigl(G_{i-j}\prodl_lH_l^{\e li}\bigr)\id,n$.
It is obvious that we can choose this \esp/ $\H^0$ such that
$t_i^0\vst{{\H^0}}=z_i^0$. So an \irc/ $\pi^0\st\pi\vst{{\H^0}}$ is an \Airp/
such that: $\Si^{\pi^0}=\Si$.
\EPf
\Pf of Theorem~2.
This Theorem simply follows from the formula (3.11) and the Theorem~1\,.
Let $\pi_0$ be an \Airp/, $\Si=\Si^{\pi_0}$ and the \rp/ $\pi=\pi\usM$ is built
as described above. One can see that operators $t_i^0$ are organized as
products of commuting factors
$t_{ik}=\prodT{j=kn+1}{(k+1)n}\xi_j\bigl(G_{i-j}\prodl_lH_l^{\e li}\bigr)$.
Let $\H^k$ be a \mcs/ of $t_{ik}\id,n$ and
$\bigotimes\limits_{k=1}^n\H^k\st\H^0$. Taking $\pi^k$ as an \irc/ of
$\pi(\si_{kn+1},\ld,\si_{(k+1)n})\vst{{\H^k}}$ it is easy to see that
$\pi^k$ is an \frp/. The \rp/
\par\kern-6pt
$$
\pi^0=\pi^M\xld\pi^1
\FTg\gta
$$
\par\kern-2pt\nin
is an \Arp/ of degree $M$, $\dim\pi^0=\NM$ and $\Si^{\pi^0}=\Si$.
Therefore it should be \ir/, \eqt/ to $\pi_0$ and (\gta) is its decomposition
to a \tp/ of \frp/s.
\EPf
\ST \Uc\ccm/ \rp/s and \itw/s/ST
\Df
\Uc\rp/s $\pii$ of \alg/ $\A$ are called \crp/ if the \rp/s
$\pit$ and $\pti$ are \eqt/. A linear invertible operator $\R$ such that
$$
\R\pit(\Delta(\O))=\pti(\Delta(\O))\R
\FTg\ia
$$
for any $\O\in\A$ is called their \itw/.
\EndP
\Lm\LH
Let $\pii$ be \crp/ and all \cel/s are represented in $\pit$ by scalars. Then
$$
\left[\av T^{\pi_1}(v),\av T^{\pi_2}(v)\right]=0
\FTg\ib
$$
\EndP
\Pf.
The statement follows from the Lemmas \LC\,,\LF\,.
\EPf
\Lm\LI
Let $\pii$ be \Airp/s and both $\pit$ and $\pti$ be \Arp/s.
Then $\pi_1$ and $\pi_2$ cocommute if and only if Eq.(\ib) is satisfied and
their \itw/ is unique modulo scalar factor.
\EndP
\Pf.
Due to the Theorem~1 both $\pit$ and $\pti$ are \Airp/s because of their \dm/s.
So the part ``only if'' follows from the previous Lemma. On the other hand if
Eq.(\ib) is satisfied it follows from Eq.(\fcd),\(\fe) and the Lemmas {\LE}\,,
\alb{\LF} that $\Si^\pit=\Si^\pti$. Hence returning to the
Theorem~1 we obtain that they are \eqt/ \irp/s.
\EPf
So we reduce the problem to consideration of matrix \Apn/s instead of \Airp/s.
 For $\T(v)\in A\M[v]$ let $\Ml\st\M[v]$ be spanned by
$v^k\T^l(v)\,,\ \,k,l\ge0$.
\Lm\LI
Let $\Pc(v)\,,\T(v)\in A\M[v]$ and $[\Pc(v)\,,\,\T(v)]=0$. Then for
generic $\T(v)\ \ \Pc(v)\in\Ml$.
\EndP
\Lm\LJ
Let $\T(v)\in A\M[v]$ and $\Tl(v)=\T(v)-\la I$. Then for generic
$\T(v)\ \ \crk\Tl(v)\le1$ for all $\la,u$.
\EndP
\Pf.
If $\la_0,v_0$ such that $\crk \T_{\la_0}(v_0)>1$ exist then $\la_0$ is a
common zero of $\det \Tl(v_0)\,,A_{n-1}^{\Tl}(v_0)$ and $B_{n-1}^{\Tl}(v_0)$
as \pln/s on $\la$. Therefore, $v_0$ is a common zero of three their mutual
resultants as \pln/s on $v$. But it is impossible for generic $\T(v)$.
\EPf
\Pf of Lemma {\LI}. Let us recall that if $\X\in\M$ has a ``simple''
spectrum in a sense that $\crk(\X-\la I)\le1$ for all $\la$ then the set
$\{\X^k\}_{k=0}^{n-1}$ is a basis of its \cz/. A generic $\T(v)$ has a
``simple'' spectrum for all $v$, so
$\Pc(v)=\sum\limits_{k=0}^{n-1}P_k(v)\T^k(v)$. Treating this equality as a
system of linear equations for functions $P_k(v)$ we see that it
has a unique solution for any finite $v$. Taking into account the Cramer's
formulae one can see that $P_k(v)$ must be whole rational functions,
{\fsp i.e. \pln/s}. The same idea applied to the highest order terms
(infinite $v$) gives the equality for degrees:
$\deg\Pc=\max\limits_k\,(\deg P_k\,,k\deg\T)$.
\EPf
\nin Certainly, if $\Pc_1(v),\Pc_2(v)\in\Ml$ then
$[\Pc_1(v)\,,\Pc_2(v)]=0$. And \vv/, one can say that if
$[\Pc_1(v)\,,\Pc_2(v)]=0$ then \gnl/ $\Pc_1(v),\Pc_2(v)\in\Ml$ for some
$\T(v)$.
\par
Later we shall use the following trivial idea: {\it A nonzero meromorphic
function is not zero at generic point.}
\Lm\LK
 For generic \Apn/ $\Pc(v)$ its power $\Pc^m(v)$ is also an \Apn/.
\EndP
\Cor For generic $\T(v)\in A\M[v]$ and
$\Pc_1(v),\Pc_2(v)\in\Ml\quad\amb\Pc_1(v),\amb\Pc_2(v)\in A\M[v]$.
\par
Now let us return to the \itw/s. Due to the Theorem 1 the space of \Airp/s of
degree $M$ is $\UM$ and all of them can be realized in the same space
$V^M=\C^\NM$. Define $\Rc_\T$ as a set of \irp/s $\pi$ such that
$\av T^\pi(v)\in\Ml$. We want to treat an \itw/ as a function of the
intertwining \rp/s and it can be done. According to Lemmas {\LI,\LK}\, \itw/s
for \ccm/ \Airp/s of degrees $M\,,M'$ define modulo scalar factor a locally
holomorphic {\ns$\text{Hom}\,(V^M\!,V^{M'})$}-valued function on
$\bcup_\T\Rc_\T^{\times2}\cap(\UM\times\Ups_{M'})$. Moreover, this function
evidently is nearly meromorphic function, only a common scalar factor can be
multivalued. Later we imply an \itw/ to be considered as a function of \rp/s in
the sense described above.
\Lm\LKA
Let $\R(\pii)$ be an \itw/ for \ccm/ \irp/s $\pi_1\,,\pi_2$. Then \gnl/
$\tr\R(\pii)\ne0$.
\EndP
\Pf.
It is sufficient to take $\pi_1=\pi\ot l;$ and $\pi_2=\pi\ot m;$ for some
\Airp/ $\pi$ and integers $l,m$. \Uc\gnl/ $\pi\ot2;$ is also an \Airp/ and
$\R(\pi,\pi)$ is proportional to the permutation operator. Now
one can give the explicit expression for the \itw/ $\R(\pii)$ and show that
$\tr\R(\pii)\propto N^k$ where $k$ is the maximal common factor of $l$ and $m$.
\EPf
\nin This Lemma shows that $\tr\R(\pii)=1$ is a good normalization
condition making an \itw/ a pure meromorphic function.
\Lm\LL
Let $\pi_a\in\Rc_\T$, $a=1,2,3$ be \Airp/s such that all
$\pi_a\x\pi_b$ $(a\ne b)$ are \Arp/s. Then \itw/s
$\R(\pi_a,\pi_b)$ satisfy the \YB/
$$
\RA\RB\RC=\RC\RB\RA\,.
$$
\EndP
\Pf.
We consider the both sides of this equality as functions on
$\bcup_\T\Rc_\T^{\times3}$.
\par\kern-4pt
\nin Put $\Re=$
\par\kern-20pt
$$
(\RC\RB\RA)^{-1}\RA\RB\RC\ .
$$
\par\kern-4pt
\nin$\Re$ commutes with all operators of the \rp/ $\pit\x\pi_3$ which is \gnl/
an \Arp/, hence $\Re$ is a scalar. Moreover, from
$$
\RA\RB\RC\RA^{-1}=\Re\,\RC\RB
$$
we see that $\tr(\RB\RC)=\Re\,\tr(\RC\RB)$. So \alb$\Re=1$.
\EPf
\Pf of Lemma \LK. It is enough for any degree $l$ and power $m$
to give an example of a \pln/ $\Pc(v)\,,\ \deg\Pc=l$ satisfying items 1,2 of
the Definition {\DA} such that $A_k^{\Pc^m}(v)$ has simple zeros and to give
an example of a similar \pln/ $\Sc(v)\,,\ \deg\Sc=l$ such that $A_k^{\Sc^m}(v)$
and $B_k^{\Sc^m}(v)$ have no common zeros. We shall take $\Pc(v)$ as follows:
$$
\gather
\gathered
\Pc_{ii}(v)=(v-w_i)^l\qc\qq w_i\ne w_j\quad\text{if}\ \,i\ne j\\
\Pc_{i,k+1}(v)=v\qc\qq \Pc_{k+1,i}(v)=\veps
\endgathered
\qq\il,k\\
\nn3pt>
\Pc_{ii}(v)=1\qc\qq\!\!i=k+1,\ld,n\qc\qq \Pc_{ij}(v)=0\qq\text{otherwise}\qp
\endgather
$$
One can calculate that for $\veps\to0$
$$
A_k^{\Pc^m}(v)=\prod_{i=1}^k(v-w_i)^{lm}+\veps v\sum_{i=1}^k
\prod_{\ssize{j=1}\atop\ssize{j\ne i}}^k(u-w_j)^{lm}
\sum_{s=0}^{m-2}(v-w_i)^s
+o(\veps)
$$
so $A_k^{\Pc^m}(v)$ have simple zeros for small enough $\veps$.
\par
We shall seek for a \pln/ $\Sc(v)$ of the following type:
$$
\Sc(v)=\pmatrix
\ \am(v)\ &v\bb&0\\0&(w-v)^l&0\\0&0&(w-v)^lI\endpmatrix
$$
where $\am(v)$ is a {\ns$\,k$ by $k\,$} block, $\bb$ is a {\ns$k$}-column and
$I$ is the $(n-k-1)$ \dm/al unit matrix. Let $\am(v)$ be a {\ns$k$}-\dm/al
\Apn/ of degree $l$, $\det\am(v)$ has simple zeros, $\det\am(w)\ne0$ and the
\prl/ {$(k-1)$\kern-\ms-\nobreak th} minor of $\am(v)$ is not zero at zeros of
$\det\am(v)$. One can build such a matrix $\am(v)$ in a way similar to the
formulae (\gsa),\(\gsb). Let us also take $\bb\notin\I\am(v)$ at zeros of
$\det\am(v)$. The technical exercise is to show that $B_k^{\Sc^m}(v)$ is not
zero at zeros of $A_k^{\Sc^m}(v)$.
\EPf
\ST sl(n) \cPm//ST
Unfortunately, no reasonable explicit expression for \itw/s of generic \Arp/s
can be obtained directly, even for the $sl(2)$ case.
The way to obtain such an expression in this case is to use the factorization
of \Arp/s to \srp/s. As a result formulae for \itw/s through the \Bw/s of the
\cPm/ can be got [\TARJ,\BS]. The first generalization of the \cPm/ to the
$sl(n)$ case was proposed in [\BMKS,\JMM] and corresponding formulae for \itw/s
of minimal cyclic \rp/s were written. Here we will discuss another
generalization, concerning \Arp/s. Although minimal \rp/s are not \Arp/s
the same \Bw/s as in [\BMKS,\JMM] happen to be used (cf.[\KAMA]).
\par
Let us take a two-\dm/al subspace $\Pi\st\C^{2n}$ and introduce a couple
$(\Gamma,\Phi)$ where $\Gamma$ is a variety:
$$
\gather
\nn-6pt>
\Gamma=\left\{p\in\C^{2n}\,|\ \av p\in\Pi\right\}\qc\\
\nn6pt>
p=\point a_1;a_n;b_1;b_n;\qc\qq\av p=\point a_1^N;a_n^N;b_1^N;b_n^N;\qc
\endgather
$$
and $\Phi$ is an $n\!$ by $\!n$ matrix such that
\vadjust{\kern2pt}
$\dsize\Phi_{ij}^N\Phi_{kl}^{-N}={\der(a_i^N,b_j^N)\over\der(a_k^N,b_l^N)}\ $.
Here the right hand side is a Jacobian calculated on the subspace $\Pi$.
Always later we shall refer only to $\Gamma$ implying the couple
$(\Gamma,\Phi)$. Let
\vadjust{\kern2pt}
$\W$ be the quotient of \alg/ $\Wc$ modulo \rl/s $\,\F\G^{-1}=1\,$,
\vadjust{\kern-4pt}
$$
 F_i^N=1\ \ ,\ \quad G_i^N=\om 1i^{\vp{-1}}\om i1^{-1}\ \ ,\ \quad
H_i^N=1\ \ ,\qq\il,n
\vadjust{\kern-5pt}
$$
and $\Z$ be the center of $\W$.
We shall retain the same notations for generators in case of $\W$ keeping in
mind new extra \rl/s. One can see that $\Z$ is \gb/
$f_i=F_i\prodl_lH_l^{-\e il}\id,n$.
\par
Define the simplest \Lop/
$L(u,p)\in\M\x\W$ as follows:
$$
L_{ij}(u,p)=\ab i^{-1}\left(-ua_iF_i\dl ij+(-u)^\tij b_iG_i\dl{i+1,}j\right)
\qp\Bm
\FTg\ic
$$
Try to find a solution of the ``skew intertwining'' \rl/
$$
\gather
\gathered
S\pp L_2(u,\pe^1)L_1(u,p)=L_2(u,p^1)L_1(u,\pe)S\pp\qc\\
\nn6pt>
[S\pp\,,\,H_i\x H_i]=0
\endgathered
\FTg\ida\\
\nn6pt>
\line{\text{where} $S\pp\in\W\ot2;$,\hfil}\\
\nn6pt>
p^1=\points a_1;a_{n-1};a_n;b_2;b_n;b_1;\qc\qq
\pe^1=\points\a_1;\a_{n-1};\a_n;\be_2;\be_n;\be_1;
\endgather
$$
and subscripts indicate embeddings $\W\st\W\ot2;$ as \crr/ing
factors. Introduce elements
$$
J_i=F_{i+1}^{-1}G_i\x G_i^{-1}F_i\qc\qq
K_i=(H_{i+1}^{-1}\x H_i)J_i
$$
such that $J_i^N=K_i^N=(-1)^{N-1}$ and define \sub/ $\J_m\st\W\tm$ \gb/
\vadjust{\kern-7pt}
$$
J_i(k)=1\ot(k-1);\x J_i\x1\ot(m-k-1);\qq
\aligned
i&=1,\ld,n\\k&=1,\ld,m-1
\endaligned\qp
\vadjust{\kern-4pt}
$$
Define also \sub/ $\K_m\st\W\tm$ \gb/ $\Z\tm$ and
$$
K_i(k)=1\ot(k-1);\x K_i\x1\ot(m-k-1);\qq
\aligned
i&=1,\ld,n\\k&=1,\ld,m-1
\endaligned\qp
$$
\Lm\LMA
Let $p,\pe$ belong to the same variety $\Gamma$ and $\av p\ne\av\pe$.
Then there exists \gnl/ a unique modulo $\K_2$ solution $S\pp$ of Eq.(\ida):
$$
S\pp=\sum_{\s\in\bold Z_N^n} W_{p\pe}(\s)
\omega^{s_1s_n}\prod_i\omega^{(1-s_i)s_i/2}J_1^{s_1}\ldots J_n^{s_n}
\FTg\ie
$$
where
$$
\gather
W_{p\pe}(\s)=\left({\ab i^N\over\baN-\abN}\right)
^{\tsize{s_n-s_0}\over\ssize N}
\prod_i\prod_{j=1}^{s_i-s_{i-1}}{\bat\omega-\abt\omega^j\over\ab i}\ \ ,\,\,
\FTg\iea\\
\nn3pt>
s_{i-1}\le s_i\qc\qq\il,n\qc\qq s_0=s_n\pmod N\qp
\endgather
$$
\EndP
\nin{\it Note.\/} The first ratio in the r.h.s. of Eq.(\iea) actually does
not depend on $i$. Inequalities there describe a \cnv/ choice of
the representative for $\s$.
\Lm\LMB
$S\pp$ satisfies the \inr/
$$
S\pp S(\pe,p)=\Im\pp
$$
and the skew \YB/:
$$
\aligned
&(S(\pe,\pee)\x1)(1\x S(p^1,\pee^1))(S\pp\x1)=\\
&=\varrho\ppp(1\x S\pp)(S(p^1,\pee^1)\x1)(1\x S(\pe,\pee))
\endaligned
$$
where $\varrho\ppp$ is a nonzero scalar and
$$
\Im\pp=N^{n+1}\prod_i{\bat-\abt\over\baN-\abN}\cdot
{\prodl_i\baN-\prodl_i\abN\over\prodi_i\bat-\prodi_i\bat}\qp
$$
\EndP
\nin This Lemma \crr/s to the Theorem 4.1 from [\Rf\DJMM] and the \inr/
(A.1) from [\Rf\KMN]. (It should be noted that for $p\in\Gamma$ we suppose that
$p^1\in\Gamma^1$ with $\Phi^1_{ij}=\Phi_{i,j+1}$.)
\par
Introducing the products $\L^m(u,\p)$ and $\Sr(\p,\pbe)$:
$$
\align
\L^m(u,\p)&=L_m(u,p_m^m)\cld L_1(u,p_1^1)\in\M\x\Wm\qc\\
\p&=(p_1\ld p_m)\in\Gamma^{\times m}\qc
\endalign
$$
$$
\Sr(\p,\pbe)=\prod_i\,\prod_{j=i+1}^{i+n}S(p_i^i,\pe_{j-i}^{j-1})
\in\W\ot 2m;
$$
($i$ is increasing and $j$ is decreasing from left to right in this product),
we get usual intertwining \rl/
$$
\Sr(\p_1,\p_2)\L_2^n(u,\p_2)\L_1^n(u,\p_1)=
\L_1^n(u,\p_1)\L_2^n(u,\p_2)\Sr(\p_1,\p_2)
\FTg\iha
$$
where subscripts indicate embeddings
$\W\ot n;\st\W\ot n;\x\W\ot n;$.
\Lm\LMC
$\Sr(\p_1,\p_2)$ satisfies the \YB/
$$
\aligned
&(\Sr(\p_2,\p_3)\x\1)(\1\x \Sr(p_1,\p_3))(\Sr(\p_1\p_2)\x\1)=\\
&=(\Sr(\p_1\p_2)\x\1)(\1\x \Sr(p_1,\p_3))(\Sr(\p_2,\p_3)\x\1)
\endaligned
$$
where $\1=1\ot n;$.
\EndP
To prove announced Lemmas we have to study some extra subalgebras.
\Lm\LN
Let us consider \sub/ $\Sc\st\W$ \gb/ $F_{i+1}^{-1}G_i\id,n$. Then
$\Sc'$ -- the \cz/ of $\Sc$ is \gb/ $F_i^{-1}G_i\id,n$ and $\Z$.
\EndP
\Pf.
\Uc\crl/ in $\W$ are homogeneous, so modulo factors belonging to $\Z$ we have
to test only monomials
of {\ns$H_i$'s and $G_i$'s}. But $E=\prodl_iH_i^{\mu_i}G_i^{\nu_i}\in\Sc'$ if
and only if $\mu_{i+1}-\mu_i=\sum\limits_j\nu_j(\e ij-\e{i+1,}j)$, so
$E\in\prodl_i(F_i^{-1}G_i)^{\nu_i}\,\Z$.
\EPf
\Lm\LNA
The \cz/ of \sub/ $\Lo_m\st\Wm$ \gb/
$\{H_i\tm,\allowmathbreak
 F_i\ot(m-k);\x G_{i-1}\xld G_{i-k}\}_{i=1}^n\vp\}_{\,k=1}^{\,m}$
is equal to $\K_m$.
\EndP
\Pf.
Denote the \cz/ of $\Lo_m$ by $\Lc'_m$. One can check that
$\K_m\st\Lc'_m$. Obviuosly, $1\ot(m-1);\x F_{i+1}^{-1}G_i\in\Lo_m$ so
$\Lc'_m\st\W\ot(m-1);\x\Sc'$. This imply that $\Lc'_m$ is \gb/
$\Lc'_{m-1}\x1$ and $1\ot(m-2);\x\K_2$. Step by step we can reduce the problem
to $m=1$ and show that $\Lc'_m$ is \gb/ $\Lc'_1\x1\ot(m-1);$ and $\K_m$. But
$\Lo_1=\W$ and $\Lc'_1=\Z$.
\EPf
\Lm\LMD
Let $\Lp\st\Wm$ be \sub/ \gb/ $H_i\tm$, $\il,n$ and all entries of
$\L^m(u,\p)$. The \cz/ of $\Lp$ is equal to $\K_m$ for generic $\p$.
\EndP
\Pf.
One can check that $\K_m$ commute with $\Lp$. So it is enough to prove the
statement only for one variety $\Gamma$ and one point
$\p\in\Gamma^{\times m}$. We shall
use the trick of the ``\tri/ limit''[\DJMM]. Let us take $\Gamma$
containing $\dsize\po=\left({1\ldots1\atop0\ldots0}\right)$ and tend
$p_i\to\po\id,m$ one after another. In this limit $\Lp$ goes to $\Lo_m$ which
\cz/ is equal to $\K_m$ according to the previous Lemma.
\EPf
\Pf of Lemma \LMA.
Substituting the expressions (\ie) into Eq.(\ida) we get identities
$$
[S\pp\,,\,G_{i+1}\x G_i]=[S\pp\,,\,H_i\x H_i]=0
$$
and equations
$$
\align
&S\pp\FG i\left({b_i\a_i\over\ab i}J_i+{a_{i+1}\be_{i+1}\over\ab{i+1}}\right)=
\\
&=\FG i\left({a_i\be_i\over\ab i}J_i+{b_{i+1}\a_{i+1}\over\ab{i+1}}\right)S\pp
\endalign
$$
which together with \crl/
$$
J_iJ_j=J_jJ_i\omega^{\dl{i+1,}j-\dl i{,j+1}}\qc\qq
J_i(\FG j)=(\FG j)J_i\omega^{\dl i{,j+1}-\dl ij}
$$
lead to functional equations for $W_{p\pe}(\s)$:
$$
\align
{W_{p\pe}(\s)\over W_{p\pe}(\s-\bold e_i)}&=
{\Phi_{i+1,i+1}(\omega\bat-\abt\omega^{s_i-s_{i-1}})\over
\ab i(\omega b_{i+1}\a_{i+1}-a_{i+1}\be_{i+1}\omega^{s_{i+1}-s_i+1})}\qc\\
\nn6pt>
\bold e_i&=(0,\ld,\un{i\text{\rm-th}}1,\ld,0)\qp
\endalign
$$
The formula (\iea) gives a
solution of these equations. Clearly $S(p,p)=1\x1$ so $S\pp$ is \gnl/
invertible. If $\hat S\pp$ is another solution of Eq.(\ida) then the ratio
$S^{-1}\pp\hat S\pp$ commutes with $\Lc_2(\pp)$ and, hence, \gnl/ belongs to
$\K_2$.
\EPf
\Lm\LO
The intersection $\J_m\cap\K_m$ is \gb/ scalars.
\EndP
\Pf.
It is easy to see that $\J_m\cap\K_m\st\Z\tm$, but it is also clear that
$\J_m\cap\Z\tm$ is \gb/ scalars.
\EPf
\Pf of Lemma \LMB.
$\!\!\Im\pp$ commutes with $\Lc_2\pp$, so \gnl/ $\Im\pp\in\J_2\cap\K_2$ and
hence is a scalar. Therefore $S^{-1}\pp\in\J_2$ and we see that $\varrho\ppp$
commutes with $\Lc_3\ppp$ which follows to $\varrho\ppp\in\J_3\cap\K_3$. The
explicit formula for $\Im\pp$ can be obtained in the same way as the \inr/
(A.1) from [\KMN].
\EPf
\Pf of Lemma \LMC.
Consider the ratio
$$
\align
\Rep=
\bigl((\Sr(\p_2,\p_3)\x\1)(\1\x\Sr(p_1,\p_3))(\Sr(\p_1\p_2)\x\1)\bigr)&^{-1}\\
\times(\Sr(\p_1\p_2)\x\1)(\1\x \Sr(p_1,\p_3))(\Sr(\p_2,\p_3)\x\1)&\qp
\endalign
$$
Similar to the previous proof $\Rep\in\J_{3m}\cap\K_{3m}$ and is a scalar. So
it is represented by the same scalar in any \rp/ of $\W\ot 3m;$. Let $\si$ be a
nonzero \rp/ of $\W$. Taking the \rp/ $\si\ot 3m;$ of $\W\ot 3m;$ and computing
$\det\si\ot 3m;(\Rep)=1$ we see that $\Rep$ is a root of 1. Hence it is
constant. In conclusion, it is clear that $\Re(\p,\p,\p)=1$ if $\p=(p,\ld,p)$.
\EPf
Similar to (\gqa) the mapping $\phip:\A\to\Wm$:
$$
T(u)\,\lto\,(-u)^{1-m}\L^m(u,\p)\qc\qq
H_l\,\lto\,H_l\tm
$$
is a \hm/ of algebras. Let $\Wo$ be the quotient of \alg/ $\W$ over
\rl/s $f_i=1\id,n$ and $\iota:\W\to\Wo$ be the canonical projection.
One can check that $\Wo$ is a simple algebra isomorphic to
$(\E\C^N)\ot(n-1);$. Let $\so$ be the \irp/ of $\Wo$,
$\si=\so\circ\iota$ and consider the \rp/ $\pip=\si\tm\circ\phip$
of \alg/ $\A$.
\Lm\LP
$\pip$ is \crd/ for generic $\p$.
\EndP
\Pf.
It is clear that any the \irp/ of $\Wo$ can be obtained from the
construction of the Lemma {\LFC} by proper choosing of parameters. In
particular it means that all generators of $\Wo$ are represented in $\so$ by
unitary operators and the same is the fact for generators of $\Lo_m$ in the
\rp/ $\si\tm$ modulo scalar factors. Hence $\si\tm$ is \crd/ \wrt/ $\Lo_m$ and
\gnl/ \wrt/ $\Lp$. (Use ``\tri/ limit''.) Since $\I\phip=\Lp$
the statement is proved.
\EPf
\Lm\LME
\Uc\isp/s of $\pip$ are \inv/ \wrt/ $\si\tm(\J_m)$.
\EndP
\Pf.
It suffices to prove the statement only for generic $\p$ where $\pip$ is \crd/.
Moreover, we can look to only \irs/s. Let $P$ be projector onto such subspace
along all others. As $(\so)\tm$ is the faithful \irp/ of $(\Wo)\tm$ we can
write $P=\si\tm(\O)$ with some $\O$
belonging to the \cz/ of $\Lp$ which is equal to $\K_m$ for generic $\p$.
Therefore $\O$ commute with $\J_m$ and $\I P$, $\ker P$ are \inv/ \wrt/
$\si\tm(\J_m)$
\EPf
\Cor
\Uc\isp/s of $\pip$ do not \gnl/ depend on $\p$.
\EndC
\Pf.
Let \sub/ $\Lb\st\Wm$ be \gb/ $\Lo_m$ and $\J_m$. Clearly for any
$\p\ \ \Lp\st\Lb$. Together with the Lemma it means that \isp/s of $\si\tm$
\wrt/ $\Lb$ are also \isp/s of $\pip$ for generic $\p$ and \vv/.
\EPf
\Lm\LQ
\Uc\ir/ parts of $\pi_n(\p)$ are \Airp/s for gen\-e\-ric $\p$.
\EndP
\Pf.
It is sufficient to consider only one variety $\Gamma$. Let us take it such
that $\dsize\left({a,\ld,a\atop b,\ld,b}\right)\in\Gamma$ for any $a\,,b$.
One can easily reduce the problem to the following one: To prove that \gnl/
$\U(v)=(-v)^{-1}\prodl_kU^{(k)}(v)\in A\M[v]$ where
$U_{ij}^{(k)}=-v\dl ij+(-v)^\tij b_k\dl{i+1,}j$.
Computing $\U(v)$ explicitly we can see that
$\U_{ij}(v)=(d_n-v)\dl ij+(-v)^\tij d_l\dl{i+l,}j$ where
$\prodi_k(b_k-v)=\sum_ld_lv^{n-l}$.
Taking $d_1\,,d_{n-l}\,,d_n\ne0$ and $d_l=0$ otherwise we obtain that
$A_1(v)=d_n-v\,,\,B_1(v)=-vd_1$ if $l=0$ and
$A_{l+1}=(d_n-v)^{l+1}+v^ld_1^ld_{n-l}\,,\, B_{l+1}=-vd_1(d_n-v)^l\,,\,
C_{l+1}=v^{l-1}d_1^{l-1}d_{n-l}(v-d_n)$
if $l>0$. Therefore \gnl/ $\U(v)\in A\M[v]$.
\EPf
\Uc\ir/ parts of $\pi_n(\p)$ is called \fcrp/s. Finally, we have got the
following picture. Let $V$ be an \irs/ of $\si\ot n;$ \wrt/ $\Ld_n$
and $\pi(\p,V)=\pi_n(\p)\vst V$. One can see that $\dim V=N^{n(n-1)/2}$.
The subspace $V$ suffices to collect all \fcrp/s because for any $\p$ and \irs/
$V'$ one can find $\p'$ such that
$\pi(\p,V')=\pi(\p',V)\,,\ \p,\p'\in\Gamma^{\times n}$.
Let $\Pb$ be the permutation operator \crr/ing to $\si\ot n;\x\si\ot n;$. Then
in virtue of (\iha) the \rp/s $\pi_n(\p)$ and $\pi_n(\p')$ are \ccm/ if
$\p\,,\p'$ are in the same variety $\Gamma^{\times n}$ and
$\R(\p,\p')=\Pb\si\ot2n;(\Sr(\p,\p'))$
is their \itw/ in the sence of Eq.(\ia). $\R(\p,\p')$ can be restricted to
$V\x V$ giving the \itw/ for \ccm/ \fcrp/s $\pi(\p,V)\,,\pi(\p',V)$. So we
have got an explicit formula for an \itw/ of special \frp/s -- \fcrp/s.
Unfortunately, counting of parameters shows that \fcrp/s do not cover the
total set
of \frp/s. On the other hand it is not surprising because we can see from the
Lemma {\LI} that a generic variety of \ccm/ \frp/s is 3-\dm/al but a variety
of \ccm/ \fcrp/s is at least {\ns$(n+1)$-}\dm/al, which is larger for $n>2$.
\ST \Uc\qm/s and \qdet//ST
Now we want to discuss some technical problems skipped before. In this section
it is not necessary to suppose that  $\e ij$ are integers and $\omega$ is a
root of 1. Only the condition $\om ij\om ji=\omega^{1+\dl ij}$ is assumed.
It is more \cnv/
to study a little bit more general situation. We introduce a new \Rm/ $\Rb(u)$
replacing in Eq.(\fa) a tensor $\eps$ by a similar tensor $\beps$ and change
the definition of \alg/ $\A$ substituting $\Rb(u)$ instead of $R(u)$ in the
left hand side of the \rl/ (\fca):
$$
\Rb(u)\ov1T(uv)\ov2T(v)=\ov2T(v)\ov1T(uv)R(u)\qp
\FTg\ka
$$
Let $V=\Cn$ and $e_1,\ld,e_n$ be the canonical basis of $V$.
Later we regard \mm/ as matrices over $\A$, naturally acting in the \Abm/
$V_\A=\A\xc V$. We assume the embedding $1\x\text{id}:V\to V_\A$ taking
place. Let us introduce the \Abm/s
$V\tm=\underbrace{V_\A\xa\ldots\xa V_\A}_m=\A\xc V\tm$ and their \sbm/s
$\VA m;=\A\xc\V m;\!$, $\V m;$ being spanned by completely antisymmetric
tensors. Define $b_m\in\E V\tm$ as follows:
$$
b_m\,e_{i_1}\xld e_{i_m}=\prod_{l=1}^m\prod_{k=1}^{l-1}
\omega_{i_ki_l}^{\theta_{i_li_k}}e_{i_1}\xld e_{i_m}\qp
$$
$\B_m$ is defined by the similar formula with $\omb ij=\omega^{\eb ij}$
instead of $\om ij$. One can check that
$$
\V m;=b_m\Big(\bigcap_{k=1}^{m-1}\ker\Rk\Big)
=\B_m\Big(\bigcap_{k=1}^{m-1}\ker\Rbk\Big)\qp
$$
As usual the definition of \qm/s is based on the \fpr/ [\Rf\KRS].
By virtue of the \rl/
$$
\Bm\Rbk\ov kT(u\omega^{1-k})\ov{k-1}T(u\omega^{-k})=
\ov{k-1}T(u\omega^{-k})\ov kT(u\omega^{1-k})\Rk
\FTg\kc
$$
following from (\ka) $\VA m;$ is an \ism/ for $\Tm$:
$$
\Tm=\B_m\ov1T(u)\cld\ov mT(u\omega^{1-m})b_m^{-1}\qp
$$
\Df
$\Te m;(u)=\Tm\vst{{\VA m;}}\ ,\quad\Td=\Te n;(u)$\newline
Entries of $\Te m;(u)$ are called \qm/s and $\Td$ is called the \qdet/.
\EndP
\Pf of Lemma \LE.
Eq.(\fcd) gives the correct coproduct only in the original case:
$R(u)=\Rb(u)\,,\,b_m=\B_m$. In this case it is obvious from the definition that
$\Delta(\Te m;(u))=\Te m;_1(u)\Te m;_2(u)$.
\EPf
\Pf of Lemma \LA.
 For a moment we have to indicate explicitly \Rms/ taking part in the relations
defining the \amm/. Three such algebras are necessary:
$\A=\A_{\Rb R}\,,\,\A_{\vp{\Rb}RR}$ and $\A_{\Rb\Rb}$. The \YB/ shows that
\Rms/ $R(u)\,,\,\Rb(u)$ generate some \rp/s $\chi,\chb$ of \alg/s
$\A_{\vp{\Rb}RR}\,,\,\A_{\Rb\Rb}$ in $\Cn$ respectively. Taking the
{\ns$m$-th} tensor power of Eq.(\ka) and using the definition of the \qdet/
we have got
$$
\gather
\Td\,\rhb(u/v)T(v)=T(v)\rho(u/v)\,\Td
\FTg\kd\\
\nn3pt>
\line{where\hfil$\rho(u)=f(u)\,(\detq T_{\vp{\Rb}RR})^\chi(u)\qc\qq
\rhb(u)=f(u)\,(\detq T_{\Rb\Rb})^{\chb}(u)$\hfil}
\endgather
$$
and $f(u)$ is an arbitrary scalar factor. The easiest way to calculate
$\rho(u)\,,\rhb(u)$ explicitly is to use different expressions for $\Td$ for
calculating different entries of $\rho\,,\rhb$. For each entry the most
conveinient expression has only one nonzero term. As a result the matrices
$\rho\,,\rhb$ can be written as follows:
$$
\rho(u)=\prodl_{k,l}\oh l^{\e kl}\qc\qq\rhb(u)=\prodl_{k,l}\oh l^{\eb kl}
$$
and using of Eq.(\fcc) in case of $R(u)=\Rb(u)$ completes the proof.
\EPf
\Cor
$[\Td\,,\detq T(v)]=0$.
\EndC
\Pf.
It follows from (\kd) since $\det\rho=\det\rhb$.
\EPf
Let us identify $V$ with $\V(m-1);$ and $\V2;$ with $\V(m-2);$ as
follows:
$$
\alignat3
\kern-18pt
&\,e_i&&\lr&&(-1)^{n-i}e_1\lld e_{i-1}\land e_{i+1}\lld e_n
\FTg\kf\\
\nn1pt>
\kern-18pt
e_i&\land e_j\ &&\lr&\ (-1)^{i+j}&
e_1\lld e_{i-1}\land e_{i+1}\lld e_{j-1}\land e_{j+1}\lld e_n\kern-17.6pt\\
\kern-18pt
&&&\kern1.5pt i<j&&
\endalignat
$$
and take the elements written above as standard basic elements for these
spaces.
\Lm\LSA
$$
\gather
T(u)\>d^{-1}(\Te(n-1);(u\omega^{-1}))^t\db=\Td\qc
\FTg\kg\\
\nn3pt>
\Te2;(u)\>\ell^{-1}(d\w2;)^{-1}(\Te(n-2);(u\omega^{-2}))^t\>\db\w2;\>\lb=\Td\qc
\FTg\kh\\
\nn2pt>
d=\prod_l\prod_{k=1}^{l-1}\oh l^{\e lk}\qc\qq
\db=\prod_l\prod_{k=1}^{l-1}\oh l^{\eb lk}\qc\\
\ell\,e_i\land e_j=\om ije_i\land e_j\qc\qq
\lb\,e_i\land e_j=\omb ije_i\land e_j\qc\qq
\ell\,,\lb\in\E\V2;\qp
\endgather
$$
\EndP
\Pf. One has the natural embeddings $\VA n;\st V_\A\xa\VA(n-1);\st V\tm$ and
$\VA n;\st\VA2;\xa\VA(n-2);\st V\tm$, so $\Td$ can be calculated in two steps.
At first $\Tm$ is restricted to the tensor product $V_\A\xa\VA(n-1);$ or
$\VA2;\xa\VA(n-2);$ and then to $\VA n;$. Taking into account \rl/s (\kf) in
this calculation we obtain the statement.
\EPf
\Cor
$$
\gather
\check R(u)\Tov1(uv)\Tov2\(v)=\Tov2(v)\Tov1(uv)\hat R(u)\qc
\FTg\kha\\
\nn3pt>
\hat R(u)=(\rho\x I)(R(u))^t(I\x\rho)^{-1}\qc\qq
\check R(u)=(\rhb\x I)(\Rb(u))^t(I\x\rhb)^{-1}\qp
\endgather
$$
\EndC
\Pf.
One can transform Eq.(\ka) to this formulae using the Lemma {\LA} and
Eq.(\kg).
\EPf
Let us also introduce $\Tt(u)$ as follows:
$$
\Tt(u)=\ell\>b_2\>\ov2\rho^{-1}(\Tov2(u\omega^{-1}))^t(\Tov1(u))^t
\>\ov2\rhb\,\B_2^{-1}\>\lb^{-1}\qp
$$
Eq.(\kha) shows that $\VA2;$ is an \ism/ for $\Tt(u)$ and one can put
$\hat T(u)=\Tt(u)\vst{{\VA2;}}$. Using Eq.(\kd)--(\kh) one can show that
$$
(\hat T(u))^t=\Td\,\Te(n-2);(u\omega^{-1})\qp
\FTg\ki
$$
\par
Due to the structure of the \Rms/ (see Eq.(\fa)) we can consider submatrices
of $T(u)$ as \mm/ of smaller size: \crl/ inside a submatrix are also described
by the \rl/ (\ka) if one substitute there for the original matrices submatrices
of $T(u)\,,\amb R(u)\,,\amb\Rb(u)$ corresponding each others. And \qm/s of
$T(u)$ are \qdet/s of its submatrices treated as smaller \mm/. This is the
important thing permitting us to compute \crl/ of \qm/s step by step by
means of Eq.(\kg),\(\kha),\(\ki).
\Lm\LSB
Let $T\ij\,,T_{\>\kbb\>\lbb}$ be \qm/s and one of them includes another.
Then
$$
\gather
T\ij(u)T_{\>\kbb\>\lbb}(u)=T_{\>\kbb\>\lbb}(u)T\ij(u)
\Psi_{\jbb\>\lbb}\Psb_{\ibb\>\kbb}^{-1}\qc\\
\Psb_{\ibb\>\kbb}=\prodi_{i\in\ibb}\prod_{k\in\kbb}\omb ik\qc\qq
\Psi_{\jbb\>\lbb}=\prodi_{j\in\jbb}\prod_{l\in\lbb}\om jl
\endgather
$$
where bold letters are multi-indeces.
\EndP
\Pf.
If the smaller minor is an entry of $T(u)$ the statement follows from the proof
of the Lemma {\LA} because the larger minor can be considered as a \qdet/.
The general case can be got simply by multiplication.
\EPf
Now we can prove \rl/s (\gba)--\(\gbd),\(\gna) for \qm/s. Some of Eq.(\gba)
and (\gbb) are evident and others follow from the Lemma~{\LSB}.
Eq.(\gbc) can be obtained from the \rl/ (\kha) applied to the \prl/ submatrix
\gb/ the first $(i+1)$ rows and columns (its \qdet/ is the \qm/
$\Ah_{i+1}(u)$). The \rl/ (\ki) applied to the same submatrix leads to
Eq.(\gbd). And the same \rl/ applied to the submatrices generating
\qm/s $\Bh_{kl}(u)$ and $\Ch_{kl}(u)$ gives Eq.(\gna)
\ST Comultiplication of \cel/s/ST
The \fpr/ is also very helpful in handling of \cel/s. Now we again require
$\omega$ to be a primitive {\ns$N$-th} root of 1 but $\e ij$ can still be
complex. Let $W^m$ be the kernel of the complete symmetrizing projector in
$V\tm$. It is clear that $\ker\Rk\st b_mW^m$ if $k<m$. Define
$$
\R^m=\Big(\prod_{j=1}^m\prod_{i=1}^{j-1}\ov{\ ij}R(\omega^{j-i})\Big)\,b_m
$$
both indices growing from right to left. $\R^m$ will be considered as function
of $\om ij$.
\par
Let $V\sav\st V\ot N;$ be the subspace
spanned by the elements $e_i\ot N;=e_i\xld e_i\,,\amb\ \,i=1,\ldots,n$ and
$V_\A\sav=\A\xc V\sav$.
\Lm\LTB
\Uc\gnl/ $\ker\R^N=W^N\oplus V\sav$.
\EndP
\Pf.
Using the \YB/ (\fca) one can move any factor $\Rk$ in the product for $\R^N$
to the very right and show that $W^N\st\ker\R^N$. It is also clear that
$\ov{\ ij}R(u)\vst{{V\sav}}=1-u\omega$. Evidently $W^N\cap V\sav=0$. So
$W^N\oplus V\sav\st\ker\R^N$ and it remains to prove that \gnl/
$\dim\ker R^N=\dim W^N+\dim V\sav$. Here the right hand side does not depend
on $\om ij$ at all and it is enough to calculate the left hand side only for
one special case. Let us test the limit $\om ij\to0$ for $i<j$. In this limit
$$
\align
&\kern28pt R(u)e_i\x e_i=(1-u\omega)e_i\x e_i\qc\\
&{\alignedat2
&R(u)e_i\x e_j&&=(1-\omega)e_j\x e_i+o(1)\\
&R(u)e_j\x e_i&&=\om ij((1-u)e_j\x e_i+o(1))
\endaligned}
\qc\qq i<j\qp
\endalign
$$
 From this equalities one can see that $\R^N_0=\lim\limits_{\om ij\to0}\R^N$ is
finite and $\I\R^N_0$ is spanned by
$\{e_{i_1}\xld e_{i_N}\!:i_1\ge\ldots\ge i_N\,,\,i_1\ne i_N\}$. Hence
$\dim\ker\R^N_0=\dim W^N+\dim V\sav$. But \gnl/
$\dim\ker\R^N\le\dim\ker\R^N_0$, so the statement is proved.
\EPf
\Lm\LTC
Let $K\in\M\ot N;$ be a projector such that $W\st\ker K$ and $V\sav\st\I K$.
Then $K\TN\vst{{W_\A}}=0$ and $K\TN\vst{{V\sav}}=\av T(u^N)$.
\EndP
\Pf.
By virtue of the Eq.(\kc) $W_\A^N$ is an \ism/ for $\TN$. Due to Eq.(\ka) one
has the \rl/
$$
\R^m\Tm=\ov mT(u\omega^{1-m})\cld\ov1T(u)\R^m\qp
$$
which shows that $\A\xc\ker\R^N$ is also an \ism/ for $\TN$. Therefore
according to the Lemma~{\LTB} $W_\A\oplus V_\A\sav$ is its \ism/ too and the
statement follows from the straightforward computation.
\EPf
\Pf of Lemma \LF. This Lemma is a \cor/ of the definition (\fcd) of the
coproduct and the previous Lemma.
\EPf
\Pf of Lemma \LC.
Let $\chi,\chb$ be the \rp/s of \alg/s $\A_{\vp{\Rb}RR}\,,\A_{\Rb\Rb}$ \gb/
\Rms/ $R(u)\,,\Rb(u)$ in $\Cn$. Eq.(\ka) and the Lemma {\LTC} together give
$$
\gather
\Rb\sav(v)\av{\ov1T}(u^Nv)\ov2T(u)=\ov2T(u)\av{\ov1T}(u^Nv)R\sav(v)\qc\\
\nn3pt>
R\sav(v)=\av T^{\chi}(v)\qc\qq\Rb\sav(v)=\av T^{\chb}(v)\qp
\endgather
$$
$R\sav(v)\,,\Rb\sav(v)$ can be calculated easily and are equal to $(1-v)I\x I$
if all $\e ij$ are integers.
\EPf
\Pf of Lemma {\LD}.
Since all entries of $\av T(v)$ mutually commute its minors can be defined as
usual. The slightly more general statement will be proved.
\par
{\nin \sl Let $T\ij(u)$ be a \qm/ of $T(u)$ and $\av T\ij(v)$ be the
\crr/ing minor of $\av T(v)$. Then
$$
\gather
\av{T\ij}(v)=\av T\ij(v)
\prod_{\ssize{i,k\in\ibb}\atop{\ssize i>k}}\bar\tau_{ik}
\prod_{\ssize{j,l\in\jbb}\atop{\ssize j>l}}\tau_{jl}\qc
\FTg\kj\\
\nn2pt>
\bar\tau_{ik}=(-1)^{(N-1)\eb ik}\qc\qq\tau_{jl}=(-1)^{(N-1)\e jl}\qp
\endgather
$$
}\par
As before we treat \qm/s of $T(u)$ as \qdet/s of its submatrices. So we have to
prove this formula only for the \qdet/ supposing that it is proved yet for all
proper \qm/s. The complete set of formulas for all \qm/s can be obtained by
induction \wrt/ the minor's size. The base of the induction is the case when
a minor is simply an entry; in this case the formula (\kj) is tautological.
In order to prove the formula (\kj) for the \qdet/ let us  take the
{\ns$N$-th} tensor power of Eq.(\kg). Using the \crl/ (\kd) to carry
$\Td$ through $T(v)$ we come to
$$
\gather
\TN\TNh=\av{\detq T}(u^N)\qc
\FTg\kk\\
\nn2pt>
\aligned
\TNh=b_N(d\>\vp{d}\ot N;)^{-1}\prodT i{}\ov i\rho^{\,(N-i)}\,(\Tov N(u))^t&\cld
(\Tov1(u\omega^{N-1}))^t\times\\
\nn-5pt>
&\times\prodT i{}\ov i{\rhb}^{\,i}\,\db\>\vp{\db}\ot N;\,\B_N^{-1}\qp
\endaligned
\endgather
$$
Let $K$ be the same projector as in the Lemma {\LTC}. By the straightforward
computation taking into account Eq.(\kj) for proper minors one can check that
$$
K\TNh\vst{{V\sav}}=
\av T\w(n-1);(u^N)\,\prodl_i\prodl_{j=1}^{i-1}\tau_{ij}\bar\tau_{ij}\qp
$$
Now Eq.(\kk) multiplied by $K$ from the left side gives the
required formula
$$
\prod_i\prod_{j=1}^{i-1}\tau_{ij}\bar\tau_{ij}\,\av{\detq T}(v)=
\av T(v)\av T\w(n-1);(v)=\det\av T(v)\qp
\FTg\kka
$$
\EPf
\Pf of Lemma \LCB.
The only nontrivial property to be checked is
$$
\det\av T^\pi(v)=\av{Q^\pi}(v)\prod_{i,l}h_l^{\e il}=\av{\detq T}^\pi(v)\qp
$$
But it was already proved above (cf.~Lemma {\LA} and Eq.(\kka)).
\EPf
\ST Algebra of \mm/ and $\Uq$./ST
Let us make two remarks about the structure of \alg/ $\A$. At first there
exists an algebra isomorphism between $\A_{\vp{\Rb}RR}$ and $\A_{\Rb\Rb}$ if
$\eb ij=\e ij+s_{ij}-s_{ji}$ for some integers $s_{ij}$. It looks as follows
$$
\align
\A_{\Rb\Rb}\ni T_{ij}(u)\,&\lto\,
\prod_lH_l^{s_{il}}\,T_{ij}(u)\,\prod_lH_l^{-s_{lj}}\in\A_{\vp{\Rb}RR}\\
\nn3pt>
H_l\,&\lto\,H_l\qp
\endalign
$$
This mapping does not preserve the coproduct  so it is not a bialgebra
isomorphism.
\par
Now let us take a \prp/ $\pi$ of degree $M$ such that
$t_i^0=t_i^\infty=1\id,n$. We put $T(u)\equiv T^\pi(u)\,,\,H_l\equiv H_l^\pi$
and
introduce operators $E_i\,,F_i\,,G_i\id,n$ as follows:
$$
\gather
E_i=(T_{ii}^0)^{-1}T_{i+1,i}^0\qc\qq
 F_i=(T_{ii}^\infty)^{-1}T_{i,i+1}^\infty\qc\\
\nn3pt>
G_i=(T_{ii}^\infty)^{-1}T_{i+1,i+1}^0=\prod_lH_l^{-(\e il+\e l{,i+1})}\qp
\endgather
$$
 For $n>2$ they satisfy \crl/
$$
\adb
\gather
H_lE_i=E_iH_l\omega^{\dl li-\dl l{,i+1}}\qc\qq
H_lF_i=F_iH_l\omega^{\dl l{,i+1}-\dl li}\qc\qq
\prod_lH_l=1\qc\\
[E_i,F_j]=(\omega-1)G_i(H_{i+1}-H_i)\dl ij\qc\\
\nn6pt>
E_iE_j=E_jE_i\omt{ij}\qc\qq
 F_iF_j=F_jF_i\omt{ji}\qc\qq|i-j|>1\qc\\
\nn6pt>
\alignedat2
\omt{ji}E_i^2E_j&-(\omega+1)E_iE_jE_i&+\omt{ij}E_jE_i^2&=0\\
\nn3pt>
\omt{ij}F_i^2F_j&-(\omega+1)F_iF_jF_i&+\omt{ji}F_jF_i^2&=0
\endalignedat
\qc\qq|i-j|=1\quad
\endgather
$$
which look similar to the \crl/ for $\Uq$. More precisely, for $\omega^N=1$,
$N$ being odd, $\e ij={N+1\over2}(1+\dl ij)$ and $q=\omega^{(N+1)/2}$ the
operators
$$
k_i=H_i^{(N+1)/2}\qc\qq e_i={E_i\over(q-q^{-1})}\qc\qq
f_i={F_i\over(1-\omega)}
$$
satisfy the \crl/ for $\Uq$ at level 0.
\STE Acknowledgements/ST
I am gratefully acknowledge to A.N.Kirillov, M.L.Nazarov and
Yu.G.Stro\-ga\-nov for helpful discussions. I also wish to thank T.Miwa and
RIMS group for warm hospitality.
\par
\newpage
\STE References/ST
%
%
\def\no#1.#2[]{\item{#1 }\ignorespaces#2\unskip.\par}
\def\jr#1/#2/#3/{{\nineit#1 \vl#2/(#3) }}
\def\book#1/{{\nineit#1 }}
\def\vl#1/{{\ninebf#1 }}
\def\LMP{Lett. Math. Phys.}
\def\TMP{Theor. Math. Phys.}
\def\CMP{Comm. Math. Phys.}
\def\DAN{Doklady AN SSSR}
\def\SMD{Soviet Math. Dokl.}
\def\IJMP{Int. J. Mod. Phys.}
\def\NP{Nucl. Phys.}
\def\FA{Func. Anal. Appl.}
%
%
\ninerm
\baselineskip=11pt
\frenchspacing
\no\TF.
L.A.Takhtajan and L.D.Faddeev, \jr Usp. Math. Nauk/34/1979/no.5 13--63;
(\jr Russian Math. Surv./34/1979/no.5 11--68)[]
\no.
E.K.Sklyanin, L.A.Takhtajan and L.D.Faddeev, \jr\TMP/40/1979/194--220[]
\no.
L.D.Faddeev, \jr Soviet Sci. Rev.,
Section C: Math. Phys. Rev./1/1980/107--155[]
\no.
A.G.Izergin and V.E.Korepin, \jr\NP/B205[FS5]/1982/401-413[]
\no.
P.P.Kulish and E.K.Sklyanin, \jr Lect. Notes Phys./151/1984/61--119[]
\no\DRI.
V.G.Drinfeld, \jr\DAN/283/1985/1060--1064; (\jr \SMD/32/1985/254--258)[]
\no\BAZH.
V.V.Bazhanov, \jr\CMP/113/1987/471--503[]
\no\JMA.
M.Jimbo, \jr\CMP/102/1986/537--548[]
\no\JM.
M.Jimbo, \jr\LMP/11/1986/247--252[]
\no\RTF.
N.Yu.Reshetikhin, L.A.Takhtajan and L.D.Faddeev,
\jr Algebra i Analis/1/1989/no.1 178--206;
(\jr Leningrad Math. J./1/1989/no.1 193--226)[]
\no.
N.Yu.Reshetikhin and M.A.Semenov-Tian-Shansky, \jr\LMP/19/1990/133--142[]
\no\BAX.
R.J.Baxter, \book Exactly solved models in statistical mechanics/
London: Academic Press 1982[]
\no\BEL.
A.A.Belavin, \jr\NP/B180[FS2]/1981/189--200[]
\no\SKLY.
E.K.Sklyanin, \jr\FA/16/1982/27--34[]
\no.
E.K.Sklyanin, \jr\FA/17/1983/273--284[]
\no\CHER.
I.V.Cherednik, \jr\TMP/43/1980/117--119[]
\no.
P.P.Kulish and E.K.Sklyanin, \jr Zap. nauch. semin. LOMI/95/1980/129--160;
(\jr J. Soviet Math./19/1982/1596--1620)[]
\no.
J.H.H.Perk and C.L.Shultz, \jr Phys. Lett./A84/1981/407--410[]
\no\TAR.
V.O.Tarasov, \jr\TMP/63/1985/175--196[]
\no\TARJ.
V.O.Tarasov, \jr\IJMP{ A}/7/1992/Suppl. 1B, 963--975;
\book Proceedings of RIMS Reseach Project (1991) ``Infinite Analysis''/[]
\no\DRII.
V.G.Drinfeld, \jr\DAN/296/1987/13--17; (\jr\SMD/36/1988/212--216)[]
\no\CHERD.
I.V.Cherednik, \jr Duke Math. J./54/1987/no.2 563--577[]
\no\KdC.
C.De Concini and V.G.Kac, \jr Colloque Dixmier//1989/471--506;
\book Progress in Math./\vl92/Birk\-h\"auser 1990[]
\no.
C.De Concini, V.G.Kac and C.Procesi, {\nineit Quantum coadjoint action}
Preprint 1991[]
\no\BS.
V.V.Bazhanov and Yu.G.Stroganov, \jr J. Stat. Phys./51/1990/799--817[]
\no\BMKS.
V.V.Bazhanov, R.M.Kashaev, V.V.Mangaseev and Yu.G.Stroganov,
\jr\CMP/138/1991/393--408[]
\no\JMM.
E.Date, M.Jimbo, K.Miki and T.Miwa, \jr\CMP/137/1991/133--147[]
\no\KAMA.
R.M.Kashaev and V.V.Mangazeev, {\nineit $N^{n(n-1)/2}\!$-State \Rm/ related
with \alb$U_q(sl(n))$ algebra at $q^{2N}=1$}, Preprint IHEP 92-32 1992
(submitted to {\nineit Mod. Phys. Lett. A})[]
\no\SKL.
E.K.Sklyanin, in \book Integrable and Superintegrable
Systems/ed. B.A.Kupershmidt, Singapore: World Scientific 1990[]
\no\DJMM.
E.Date, M.Jimbo, K.Miki and T.Miwa, \jr Pacific J. Math./154/1992/no.1 37--65[]
\no\KMN.
R.M.Kashaev, V.V.Mangazeev and T.Nakanishi, \jr \NP/B362/1991/563--585
[]
\no\KRS.
P.P.Kulish, N.Yu.Reshetikhin and E.K.Sklyanin, \jr\LMP/5/1981/393--401[]
\bye